\pdfoutput=1
\documentclass{article}

\usepackage{PRIMEarxiv}
\usepackage{multirow}
\usepackage[utf8]{inputenc} 
\usepackage[T1]{fontenc}    
\usepackage{hyperref}       
\usepackage{url}            
\usepackage{booktabs}       
\usepackage{amsfonts}       
\usepackage{nicefrac}       
\usepackage{microtype}      
\usepackage{latexsym}
\usepackage{graphicx}
\usepackage{multicol,multirow}
\usepackage{amsmath,amssymb,amsfonts}
\usepackage{mathrsfs}
\usepackage{amsthm}
\usepackage{rotating}
\usepackage{appendix}
\usepackage{ifpdf}
\usepackage[T1]{fontenc}
\usepackage{times}
\usepackage{newtxmath}
\usepackage{textcomp}%
\usepackage{xcolor}%
\usepackage{lipsum}
\usepackage{fancyhdr}       
\usepackage{graphicx}       
\graphicspath{{media/}}     

\usepackage{natbib}
\makeatletter
\newcommand{\settitle}{\@maketitle}
\makeatother

\hypersetup{
    colorlinks=true,
    linkcolor=black,
    citecolor=black,
    filecolor=black,
    urlcolor=black,
    hyperindex=true,
    bookmarks=true,
    bookmarksopen=true,
    bookmarksnumbered=true,
}

\title{Multidata Causal Discovery for Statistical Hurricane Intensity Forecasting
}

\author{
  Saranya Ganesh S., Frederick Iat-Hin Tam, Milton S. Gomez, Tom Beucler \\
  Faculty of Geosciences and Environment,\\
  Expertise Center for Climate Extremes,\\ 
  University of Lausanne \\
  Lausanne, Vaud, Switzerland\\
   \And
  Marie McGraw, Mark DeMaria, Kate Musgrave \\
  Cooperative Institute for Research in the Atmosphere, \\ 
  Colorado State University, \\
  Fort Collins, Colorado\\ 
   \And
  Jakob Runge \\ 
  Department of Computer Science,,\\
  University of Potsdam \\
  Potsdam, Germany \\
}

\begin{document}
\settitle

\begin{abstract}
Improving statistical forecasts of tropical cyclone (TC) intensity is limited by complex nonlinear interactions and difficulty in identifying relevant predictors. Conventional methods prioritize correlation or fit, often overlooking confounding variables and limiting generalizability to unseen TCs. To address this, we leverage a multidata causal discovery framework with a replicated dataset based on Statistical Hurricane Intensity Prediction Scheme (SHIPS) using ERA5 meteorological reanalysis. We conduct  experiments to identify and select predictors causally linked to TC intensity changes. We then train multiple linear regression models to compare causal feature selection with correlation, random forest feature importance, and no feature selection, across five forecast lead times from 1 to 5 days (24–120 hours). Causal feature selection consistently outperforms on unseen test cases, especially for lead times shorter than 3 days. Top causal features include vertical shear, mid-tropospheric potential vorticity and surface moisture conditions, which are physically significant yet often underutilized in TC intensity predictions. We build an extended predictor set (SHIPS$+$) by adding selected features to the standard SHIPS predictors. SHIPS$+$ yields increased short-term predictive skill at lead times of 24, 48, and 72 hours. Adding nonlinearity using a multilayer perceptron further extends skill to longer lead times, despite our framework being purely regional and not requiring global forecast data. Operational SHIPS tests confirm that three of the six added causally discovered predictors improve forecast skill, with the largest gains at longer lead times. Our results demonstrate that causal discovery improves TC intensity prediction and pave the way toward more empirical forecasts.
\end{abstract}


\paragraph{Impact Statement}: {\normalfont As data-driven tropical cyclone (TC) forecasting tools become more sophisticated, it is critical to ensure that they make accurate forecasts for the right reasons, especially for TCs not included during training. Here, we use a technique called multidata causal discovery, which identifies a unique set of predictors driving TC intensity changes over the next five days by analyzing multiple realizations of the same underlying phenomenon: TCs in the North Atlantic. Adding these predictors to an operationally used statistical TC intensity prediction scheme improves accuracy of forecasts. This shows that selecting predictors based on cause-and-effect, not just correlation, can result in improved forecasts that are more generalizable as well as interpretable, since the causally discovered predictors are physically consistent.}

\section{Introduction}
Extensive research on tropical cyclones (TCs) has advanced our understanding of key atmospheric and oceanic processes that control cyclogenesis and intensification \cite{gray1998formation,elsberry1995recent,elsberry2014advances}. However, the continued rise in coastal populations and the increasing risks of wind gusts, storm surge, extreme rainfall, and severe weather \cite{pielke1998normalized,crossett2004population} underscore the need for more accurate and robust TC intensity forecasts.  
Despite breakthroughs in short-range to subseasonal prediction \cite{camargo2019tropical, heming2019review,tang2020recent}, accurately predicting rapid TC intensity changes remains challenging \cite{demaria1999updated, wang2004current, demaria2021operational} especially beyond 24 hours, due to error growth in initial conditions, incomplete representation of key physical processes, and limitations in data assimilation \cite{emanuel2016predictability}. For example, recent National Hurricane Center verification statistics (2020--2024) indicate mean TC intensity errors of 5.1 knots ($\sim$ 2.6 m s$^{-1}$) at 12-hour lead time and 10.0 knots ($\sim$ 5.1 m s$^{-1}$) at 48-hour lead time \cite{nhc_verification_2024}. In addition to numerical weather prediction systems, statistical models play a significant role in TC forecasts by improving the skill of multi-model ensemble-based consensus forecasts \cite{demaria2022national}. Both statistical and dynamical models struggle to estimate rapid intensification (RI) as multiscale air-sea interactions and radiative-convective feedbacks leading to intensity changes are not fully understood \cite{wang2012recent}. RI events account for 20 - 30\% of intensity forecast error variance \cite{Trabing+Bell2020}, and peak intensity of rapidly intensifying TCs is routinely underestimated \cite{torn2021validation}. Although statistical schemes outperform high-resolution dynamical models in probabilistic RI prediction \cite{torn2021validation}, dynamical model intensity forecasts, including during RI events, improve substantially when advanced configurations such as storm-following nests are employed \cite{alaka2022high}.

Statistical Hurricane Intensity Prediction Scheme (SHIPS) integrates large-scale predictors from climatology, persistence, and synoptic predictors to estimate TC intensity \cite{demaria1994statistical, MarksDeMaria2003rcliper} and its forecast skill has gradually improved over time. SHIPS evolved from a ``statistical-synoptic'' to a ``statistical-dynamical'' framework through the incorporation of synoptic environmental conditions from dynamical models (e.g. GFS) \cite{demaria2022national}. The combined consensus forecast exhibits the best skill for TC intensity prediction, with SHIPS adding value to operational intensity prediction efforts \cite{cangialosi2020recent}.  
SHIPS includes an operational probabilistic RI index based on large-scale environmental predictors \cite{kaplan2010revised,kaplan2015evaluating}. Machine learning approaches have further improved 24-hour RI prediction skill, with SHIPS-compatible extensions using satellite-derived predictors, hybrid deep learning–gradient boosting frameworks, and operational consensus models \cite{su2020applying,boussioux2022hurricane,ko2023development}.

Although SHIPS incorporates predictors from ocean analyses and satellite imagery, forecast skill decreases with lead time, partly because predictors are chosen semi-empirically based on domain knowledge, climatology, and persistence \cite{demaria2005further}. To increase sample size, SHIPS regression coefficients are re-derived annually with climatological adjustments and updates to the predictor list \cite{demaria2021operational}. While the SHIPS developmental dataset provides an extensive set of environmental predictors, it may still omit key variables relevant for changes in TC intensity. 

We address this limitation by adopting a causal discovery framework \cite{runge2018causal,runge2019inferring,runge2019detecting,gerhardus2022causal,runge2023causal} to objectively identify new environmental predictors for inclusion in the operational SHIPS developmental dataset. In causal discovery, relationships among variables are represented using a causal graph, where nodes correspond to physical variables, and directed edges denote direct causal influences, potentially with time lags. While causal discovery aims to infer the full causal structure among a set of variables, causal feature selection instead uses a causal discovery algorithm to identify a subset of physically meaningful predictors that are most relevant for predicting a target variable, without reconstructing the full causal graph. Climate science studies have applied linear and nonlinear causal discovery methods \cite{ebert2012causal, runge2014quantifying, runge2019detecting} to identify spurious associations in statistical prediction models arising from common drivers or indirect associations. Once the causal graph is known, the strength of the links between the physical variables can be determined with causal inference \cite{ricard2025causal}, complementing data-driven techniques \cite{runge2023causal}. 
Despite these advances, causality is still commonly inferred using lagged correlation analysis, which does not resolve the directionality of the relationships between variables \cite{mcgraw2018memory}. As such, lagged correlation methods are susceptible to non-causal correlations from autocorrelation effects, indirect connections through a third process, or a shared driver hindering interpretability in statistical prediction models \cite{Runge2014,barnes2015impact}. Understanding the directional causality between atmospheric variables hence requires additional methodologies or causal inference approaches beyond the scope of lagged correlation analyses. 
Several studies have demonstrated the potential of causal discovery methods in tropical meteorology. For example, \cite{wijnands2016variable} used the graphical model-based structure learning approach of the Peter-Clark (PC) causal discovery algorithm to identify key variables for tropical cyclogenesis prediction and found that predictive modeling using logistic regression employing the highest ranked variables improved statistical predictive skills. Additionally, the PCMCI+ causal algorithm — which combines the PC algorithm with Momentary Conditional Independence test \cite{runge2020discovering} — has been used to identify precursor regions useful for improving seasonal TC frequency forecasts \cite{Latos_etal2024_CausalAI}. 
Motivated by these advances, we apply the PC algorithm within a multidata causal discovery framework to identify statistically significant causal links and select predictors with direct causal influence on TC intensity change for the operational SHIPS model. This approach treats individual Atlantic TCs as multiple realizations of the same underlying physical process, assuming that their intensity evolution shares a common causal structure that can be represented by a single causal graph. The multidata PC algorithm requires multivariate time series of all candidate predictors for each storm, motivating our replication of the SHIPS developmental dataset using meteorological analyses, as described in Section~\ref{sec:data}. After detailing the causal methodology and experimental design in Section~\ref{sec:Methodology}, we demonstrate in Sections~\ref{sec:Results}\ref{subsec:Results_multidatacausaldiscovery} and \ref{sec:Results}\ref{subsec:Results_causalfeatureselectionoutperforming} that causal feature selection outperforms other feature selection methods, particularly in short-range time-scales. These improvements are robust in both reanalysis-based experiments (Section~\ref{sec:Results}\ref{subsec:Results_suggestingpredictors}), where we also show that the causal predictors nonlinearly drive intensity (Section~\ref{sec:Results}\ref{subsec:Results_SHIPSvsSHIPS+}), and in operational-like settings using real-time global model analysis and forecast fields (Section~\ref{sec:Results}\ref{subsec:Results_operationaltest}). In Section~\ref{sec:Results}\ref{subsec:Results_casestudy}, we present a case study of Hurricane Larry to illustrate the physical relevance of the causal links between the selected predictors and short-term TC intensification. 

\section{Data Sources and Preprocessing\label{sec:data}}
Our analysis is based on two complementary experimental designs: (i) reanalysis-based experiments using the high-resolution ECMWF Reanalysis Version 5 (ERA5) \cite{hersbach2018era5,hersbach2020era5}, and (ii) experiments using SHIPS developmental dataset to test the added value of causal predictors compared to the operational SHIPS predictors. To define TC-centered domains and intensity change targets, we rely on the International Best Track Archive for Climate Stewardship (IBTrACS), which provides reference track and intensity data for each TC.

\subsection{IBTrACS: Reference Dataset for Intensity and Track}
IBTrACS \cite{knapp2010international} provides six-hourly best-track positions and intensity estimates. Best-track locations define TC-following domains used to extract area-averaged ERA5 predictors, while maximum sustained surface (10 m) wind speeds are used to compute intensity change targets at forecast lead times of 24, 48, 72, 96 and 120 hours (DELV24, DELV48, ..., DELV120 in m/s). To minimize potential biases related to Dvorak intensity estimates, we restrict our analysis to the North Atlantic basin, where routine aircraft reconnaissance missions provide regular in situ measurements of TC intensity. We select 247 long-lived Atlantic TCs from 2000 to 2021 with lifetime of at least four days prior to landfall.

\subsection{ERA5 and TC PRIMED Datasets}

Causal discovery is restricted to SHIPS predictors that can be replicated using ERA5; satellite-only predictors such as GOES brightness temperature and ocean heat content are excluded to ensure consistency. The ERA5 predictor inputs consist of six-hourly, area-averaged time series extracted from TC-following domains centered on IBTrACS best-track positions. Predictors are computed for multiple radial regions around the storm center to capture inner core and outer area conditions: inner-core variables are averaged over 0–2$^{\circ}$ (approximately 0–200 km), while outer-area predictors are averaged from 200–800 km or up to 1000 km, following conventions used in the extended SHIPS developmental dataset. To avoid backward-in-time causal dependencies, we only consider predictors at analysis time (00-hour), unlike operational SHIPS setup, which incorporates dynamical model forecast fields. As a result, forecast skill at longer lead times may be underestimated when synoptic-scale evolution is important. To expand the predictor set further, we use the Tropical Cyclone Precipitation, Infrared, Microwave, and Environmental Dataset (TC PRIMED) \cite{Razin2023}. TC PRIMED uses ERA5 to reconstruct operational SHIPS predictors as well as environmental and thermodynamic variables across multiple pressure levels and radial regions. An important limitation of the area-averaging approach is that it does not capture azimuthal asymmetries, which may be relevant for sheared or rapidly intensifying TCs \cite{hazelton2017analyzing,rogers2013airborne,mcneely2020unlocking} and could be considered in future experiments.

As the first set, we have the original operational predictors from the SHIPS developmental dataset (Table S1).  For causal experiments, the predictor set includes replicated SHIPS variables derived from ERA5 reanalysis fields (e.g., VMAX, POT, PER, vertical wind shear; see Table~S2 for details), directly comparable to operational baselines. This is complemented by a broader set of synoptic and thermodynamic variables sampled at multiple pressure levels and radial domains. These additional variables include divergence, vorticity, potential vorticity, equivalent potential temperature, geopotential height, relative humidity, air temperature, temperature gradients, precipitable water, and warm-core anomalies, among others. Complete variable definitions, radial averaging details, and pressure levels are provided in Tables S1–S5 of the Supplementary Information.

 A key requirement for applying causal discovery to time series is the assumption of \textit{causal stationarity}, i.e., the causal relationships between variables remain invariant over time within a given TC and across multiple storms in the multidata setting. This allows pooling of statistical evidence across time and the full training set for conditional independence testing \cite{saranya2023selecting}. Accordingly, TC time series are aligned relative to the time of minimum central pressure to satisfy the causal stationarity assumption.
The mean sea level pressure (MSLP) time series for each storm in the training set are smoothed using a Gaussian filter ($\sigma = 3\times6\mathrm {\ hours}$) to reduce noise and the time of minimum central pressure is used as the reference point for alignment. For the shorter time series, we append NaN values on either sides to make sure that the lengths are consistent with the longest time series with minimum pressure in the middle so that the evolution of different storms could be compared on a common reference frame. All predictors are subsequently standardized using the mean and standard deviation computed from the training set.
The final ERA5 dataset includes 214 predictors and the intensity-change target (DELV) at each forecast lead time. A total of 216 TCs from 2000–2019 are used within a storm-based cross-validation framework with predefined, non-overlapping training and validation splits, while an independent test set comprises 31 TCs from 2020–2021 including Hurricane Wilma (2005), retained as a particularly intense and challenging case.

\subsection{SHIPS Developmental Dataset}
To evaluate how causally informed predictors improve operational forecasting, we extract SHIPS predictors (Table S1) for the same 247 North Atlantic TC cases (2000-2021) used in the ERA5 experiments. We only used predictor values from the 00-hour initialization time in the SHIPS developmental dataset derived from GFS analysis fields, which represent the atmospheric conditions (i.e., observations assimilated into the GFS model) at forecast initialization time. This setup enables direct analogy between the SHIPS developmental dataset and the ERA5-derived dataset used for causal predictor discovery, as both rely on analysis time atmospheric fields within a purely statistical framework.
We compare the generalization skill of statistical models trained using the original 21 SHIPS predictors (``original SHIPS'') with models trained on the same predictors augmented with causally selected predictors (``SHIPS+''). These additional predictors are identified through ERA5-based causal discovery experiments but are computed from GFS analysis fields in exactly the same way as the original SHIPS predictors. This setup tests whether the predictors selected in the ERA5 replication setting improve forecast skill when integrated into the operational SHIPS framework.

\section{Methodology\label{sec:Methodology}}
\subsection{Causal Feature Selection}
Causal feature selection is performed using the open-source TiGraMITe causal discovery framework \cite{runge2018causal,runge2019detecting} which has been widely used to discover causal relationships from climate data \cite{kretschmer2016using,Kretschmer2018,samarasingheetal2021}. We employ the multidata PC (M-PC) algorithm, a modified version of the first step of the PCMCI algorithm \cite{runge2019detecting,runge2020discovering}, to discover predictors of TC intensity change while eliminating spurious statistical associations. Unlike PCMCI+, which aims to infer the full causal graph including contemporaneous causal links, this work uses only the PC algorithm for causal feature selection \cite{iglesias2024causally,runge2015optimal}. 
The ``M'' in M-PC indicates the multidata approach, where an ensemble of time series from multiple TCs is analyzed to identify a single set of causal predictors. Non-causal drivers within the ERA5 predictor set are detected and excluded through conditional independence test using linear partial correlation \cite{saranya2023selecting}. In our setup, this is done without considering time-lagged variables — each predictor is evaluated at a single timestep for the given forecast lead time. By filtering out non-causal predictors in the statistical models, this approach aims to improve the generalizability of statistical models to unseen TC cases. 

The hyperparameters of M-PC  include the minimum and maximum time lags $\left(\tau_{\mathrm{min}},\tau_{\mathrm{\max}}\right)$ and the statistical significance threshold for the partial correlation conditional independence tests $pc_{\alpha}$ which determines variable removal. Predictors are ranked according to the absolute value of the partial correlation between each predictor and TC intensity change, conditional on the remaining variables, since the partial correlation test statistic can take both positive and negative values. To assess the robustness of the selected predictors, causal discovery experiments are repeated for each forecast lead time across a wide range of statistical significance thresholds ($pc_{\alpha} \in [10^{-4},\,0.6]$; see SI for the full list) and for multiple single-lag configurations, with $\tau_{\min}=\tau_{\max}\in\{4,8,12,16,20\}$ timesteps. For each configuration, predictors exhibiting non-significant conditional dependence are excluded, yielding multiple candidate feature sets across lead times that are subsequently evaluated through cross-validation.

\subsection{Validation procedure}
Causal discovery algorithms are sensitive to hyperparameter choice \cite{lawrence2021data}, hence, the output of the causal feature selection framework should not be interpreted as truth for a complex physical problem such as TC intensity change. Instead, it may be considered as a plausible approximation of the true causal relationships. Careful cross-validation of the causal predictor lists is necessary to establish their robustness and generalizability. Following previous work \cite{saranya2023selecting}, we postulate that causally relevant predictors improve model generalization skill, whereas the inclusion of spurious or redundant features degrades performance. By testing the performance of trained models on the validation data unseen during training, our framework should identify a unique feature set that maximizes validation skill and has the highest likelihood of being directly linked to TC intensification.
We use storm-based seven-fold cross-validation on the non-test portion of the dataset. Due to year-based partitioning and the removal of storms with no valid samples after preprocessing and alignment, six folds contain 184 training storms and 32 validation storms, while one fold contains 186 training storms and 30 validation storms. Each fold yields multiple candidates of plausible causal relationships, so we train a series of multiple linear regression (MLR) models using predictor sets derived from the PC algorithm at different statistical significance thresholds ($pc_{\alpha}$). Smaller $pc_{\alpha}$ values generally yield fewer predictors, although different thresholds can result in feature sets of same cardinality. We assess the generalization capabilities of the trained MLR models with the coefficient of determination ($R^2$). 
\begin{figure*}[ht]
    \centering
    \includegraphics[width=\textwidth]{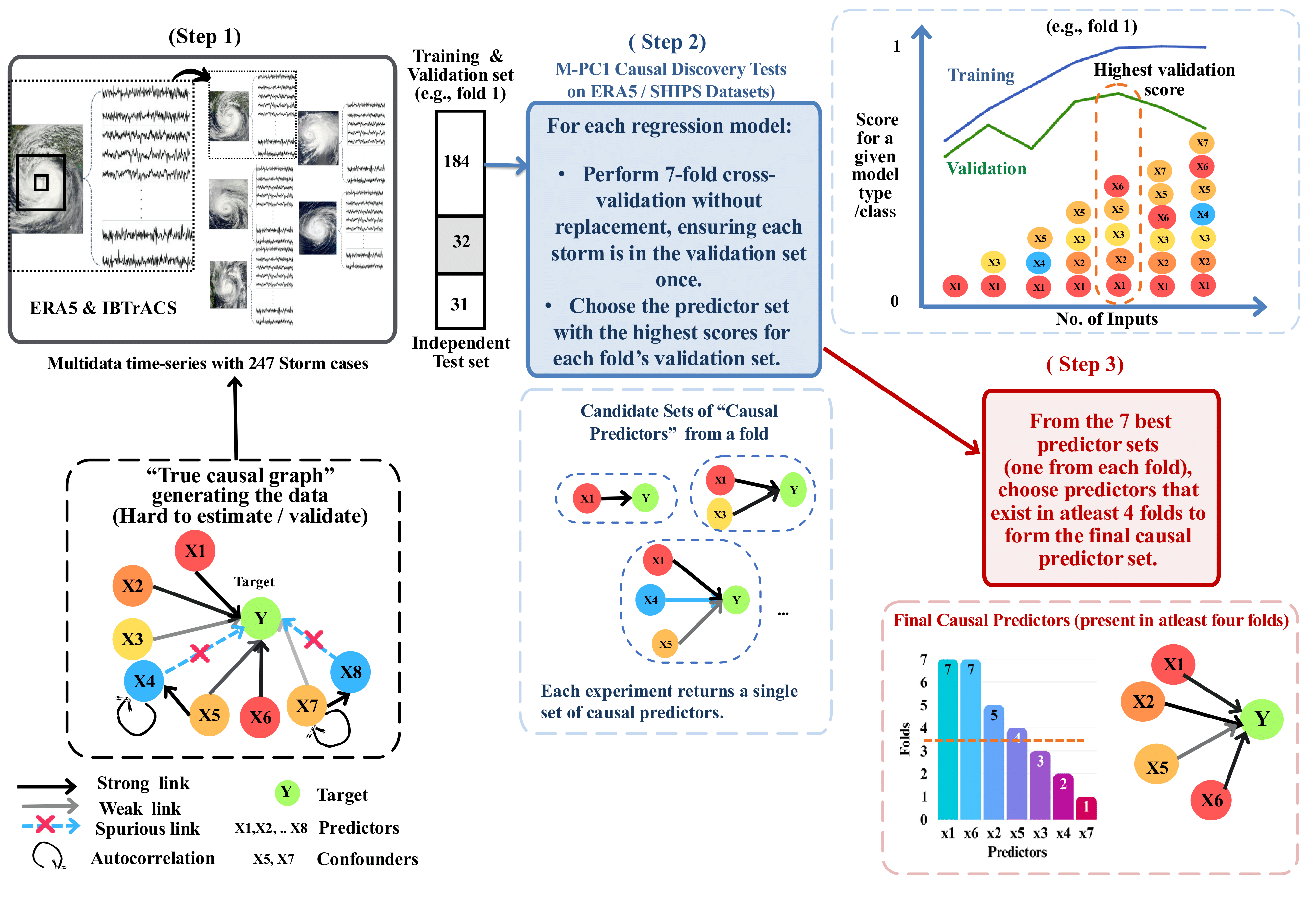}
    \caption{Multidata causal feature selection methodology. Step 1: Preprocessed spatiotemporal fields for all TC cases form an ensemble of \textit{aligned} time series, which may contain spurious or non-causal relationships due to autocorrelation or confounding. Step 2: These multivariate time series (training set) are input to the multidata causal discovery algorithm (M-PC), which selects candidate predictors while controlling set size via hyperparameters. Each candidate set is evaluated using cross-validated regression. Step 3: Predictors appearing in minimum four out of the seven folds are pooled to form the final feature set. The goal is to estimate the portion of the true causal graph that helps predict TC intensity changes.}
    \label{FIG1}
\end{figure*}
\subsection{Filtering new potential predictors}
Fig. ~\ref{FIG1} summarizes the causal feature selection and model evaluation framework. For all trained MLR models, we calculate the $R^2$ for training, validation, and test sets. For each cross-validation fold, we choose the model with the highest validation performance ($R^2$), resulting in 7 best performing models for analysis (upper-right panel in Fig. \ref{FIG1}). Comparing validation performance across all trained models allows us to identify the most generalizable MLR configuration. Adding more predictors beyond those chosen by the most generalizable MLR model will lead to model overfitting and make the MLR model less generalizable because they constitute spurious associations arising from autocorrelation, indirect effects, or common drivers. 
To ensure robustness across cases, the final causal variable list is obtained by aggregating and ranking the variables in the 7 best models. Only predictors that appear in at least four out of seven best models and are not part of the existing SHIPS developmental predictor list will be shortlisted as a candidate predictor (lower right panel in Fig. \ref{FIG1}).
\subsection{Experimental Design}
\begin{figure*}
     \centering
    \includegraphics[width=\textwidth]{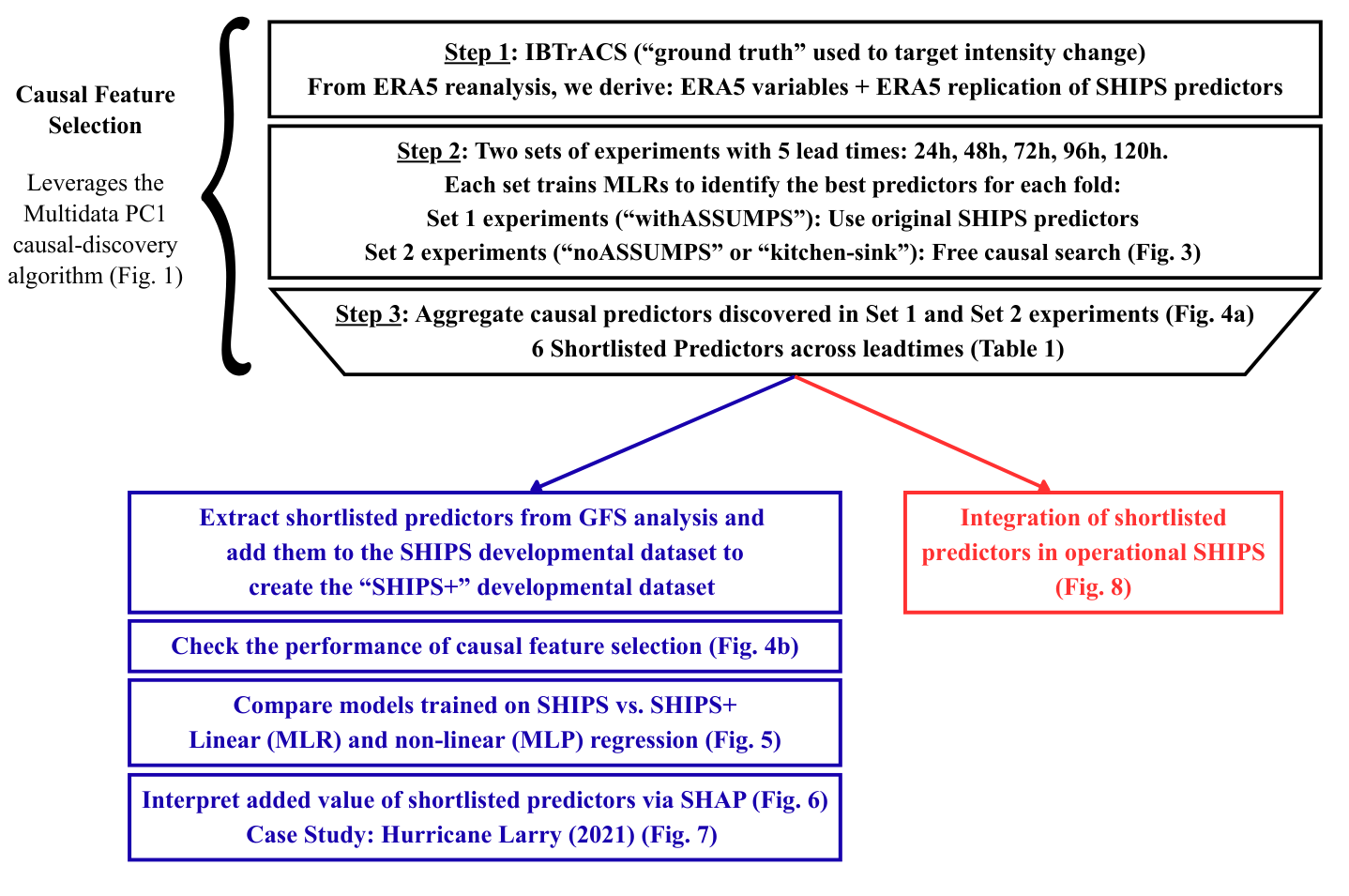}
    \caption{Workflow for causal feature selection, predictor shortlisting, and integration into SHIPS:ERA5-derived variables and ERA5-based replications of SHIPS predictors are used with IBTrACS intensity change as ground truth. Causal discovery is performed across multiple lead times under constrained (withASSUMPS) and unconstrained (noASSUMPS) setups. Predictors shortlisted across lead times based on ranking and performance are integrated into SHIPS+, followed by regression model comparison, SHAP-based interpretation, case-study analysis, and operational SHIPS testing.}
    \label{fig:overall_design}
\end{figure*}

A hierarchy of experiments is conducted to demonstrate the potential applications of causal feature selection algorithms in complementing or enhancing existing statistical models for TC intensity change. We depict our overall experimental design in Fig. ~\ref{fig:overall_design}.

\subsubsection{ERA5-based experiments for predictor discovery}
The first set of experiments involves identifying candidate causal predictors from ERA5 reanalysis data that can be incorporated into the operational SHIPS. We designed two complementary experiments to assess the algorithms and understand the sensitivity of discovered variables to the assumptions within the causal discovery process. In the first experiment (withASSUMPS), existing SHIPS predictors are retained a priori, and the causal discovery framework is restricted to identifying additional predictors. This represents a least-change and operationally feasible scenario. In contrast, the second experiment (noASSUMPS) is a free-run, ``kitchen-sink'' that allows the algorithm to freely evaluate all available predictors, including the possibility of excluding existing SHIPS predictors that do not exhibit strong causal relationships with TC intensity change.
Through this experiment, we investigate whether removing spurious associations among the original SHIPS predictors improves the generalizability of statistical intensity prediction models. 
Causal feature selection is however constrained by the properties of the underlying datasets. ERA5 does not fully resolve the inner-core structure of TCs, and limitations in representing small-scale and azimuthally asymmetric processes are therefore reflected in the candidate predictors, particularly those related to the TC inner core. As a result, some physically relevant variables may not be identified as robust causal predictors within this framework.
\subsubsection{Validation in an operation-like setting}
The second set of experiments assesses the practical value of the causal predictors identified using the ERA5 reanalysis by testing them within the operational SHIPS framework. Guided by the most robust predictors identified in the ERA5-based causal discovery experiments, we compute the corresponding predictors from GFS analysis fields and extend the SHIPS developmental dataset including these newly identified causal predictors alongside the original predictors. We then compare models trained using the original SHIPS predictor set with those trained on the augmented predictor set (SHIPS+), using the full SHIPS developmental dataset.
This comparison directly quantifies whether enriching the SHIPS developmental dataset with causally selected predictors yields measurable improvements in forecast performance on unseen TC cases.
\subsection{Regression Model hierarchy}

\subsubsection{Mapping and Optimization Objective}
Building on the operational validation setup, we predict intensity change at a fixed lead time $\tau \in \{24,48,72,96,120\}\,\mathrm{h}$, defined as
\begin{equation}
\Delta V_{\tau}=V_{\max}(t+\tau)-V_{\max}(t).
\end{equation}
Targeting $\Delta V_{\tau}$ rather than $V_{\max}(t+\tau)$ is consistent with SHIPS and our experimental design: (i) the tropical system is known to exist at initialization and $V_{\max}(t)$ is observed, so $V_{\max}(t+\tau)$ follows once $\Delta V_{\tau}$ is forecast; (ii) the target distribution is typically better behaved than the absolute intensity.

For each storm time $t$, we form a standardized predictor vector $\mathbf{x}_t \in \mathbb{R}^{p_{\tau,k}}$ composed of fold-specific causal features selected for the lead time $\tau$ (Section~\ref{sec:Methodology}). We then learn a deterministic mapping
\begin{equation}
f_{\tau,k}: \mathbf{x}_t \mapsto {\Delta V}_{\tau},
\end{equation}
with one model (and potentially one feature set) per lead time $\tau$ and cross-validation fold $k$.

Model parameters are estimated by least squares for consistency with SHIPS and with the linear, partial-correlation-based causal discovery framework. This yields a deterministic forecast without an explicit predictive distribution. We adopt this deterministic objective because (i) our focus is short lead times where mean-squared error is a standard operational target, (ii) the causal discovery framework is tailored to mean relationships rather than full conditional distributions, and (iii) maintaining compatibility with SHIPS facilitates a clean comparison in the baseline experiments. 

\subsubsection{Baseline Regression models and Baseline Feature Selection Methods}
We compare the performance of PC-based causal feature selection models to different feature selection baselines including lagged correlation and random forest feature importance. The correlation baseline is obtained by ranking the linear correlation between the input variables and rate of TC intensity change. The variable ranking is then used to sequentially train linear regression models with increasing complexity. In random-forest-based baseline, the variable ranking is done with the Gini impurity-based feature importance of trained regression models. We apply these feature selection baselines for the withASSUMPS and noASSUMPS versions across selected lead times for both ERA5 and SHIPS-based experiments.
\subsubsection{Regression architecture}
We first evaluate all predictor sets using multiple linear regression (MLR) models to preserve interpretability and isolate the impact of feature selection. For each experiment, we repeat 7-fold cross-validation across forecast lead times (24–120 hours), assessing performance on independent test sets.
We further evaluate the predictor sets using multilayer perceptron (MLP) models, which we select based on their ability to capture nonlinear behaviors in the data that are not captured by MLR models. Each of the MLP models in our study includes a total of 5 layers, wherein each layer includes 512 units (also referred to as neurons). The first three layers rely on a Rectified Linear Unit (ReLU) activation function, the fourth layer instead uses hyperbolic tangent (tanh) to allow the model to output positive and negative intensity changes, and the final layer linearly combines the output of the 4th layer. All models were trained to minimize the mean square error (MSE) using the Adam optimizer \cite{Kingma2014} with default parameters, a learning rate of 0.001, and with early stopping set to end training if the validation loss is greater than the average of the last 50 epochs once the model has trained for at least 50 epochs. This simplified architecture was chosen with rules of thumb and has not been subjected to a rigorous hyperparameter search given that the objective of these MLPs is not to achieve maximum performance, but rather to provide a simple \mbox{nonlinear} baseline for comparison to the MLRs.

\section{Results\label{sec:Results}}

\subsection{Effectiveness of multidata causal discovery for TC intensity prediction\label{subsec:Results_multidatacausaldiscovery}}
\begin{figure*}[ht]
    \centering
    \includegraphics[width=0.8\textwidth]{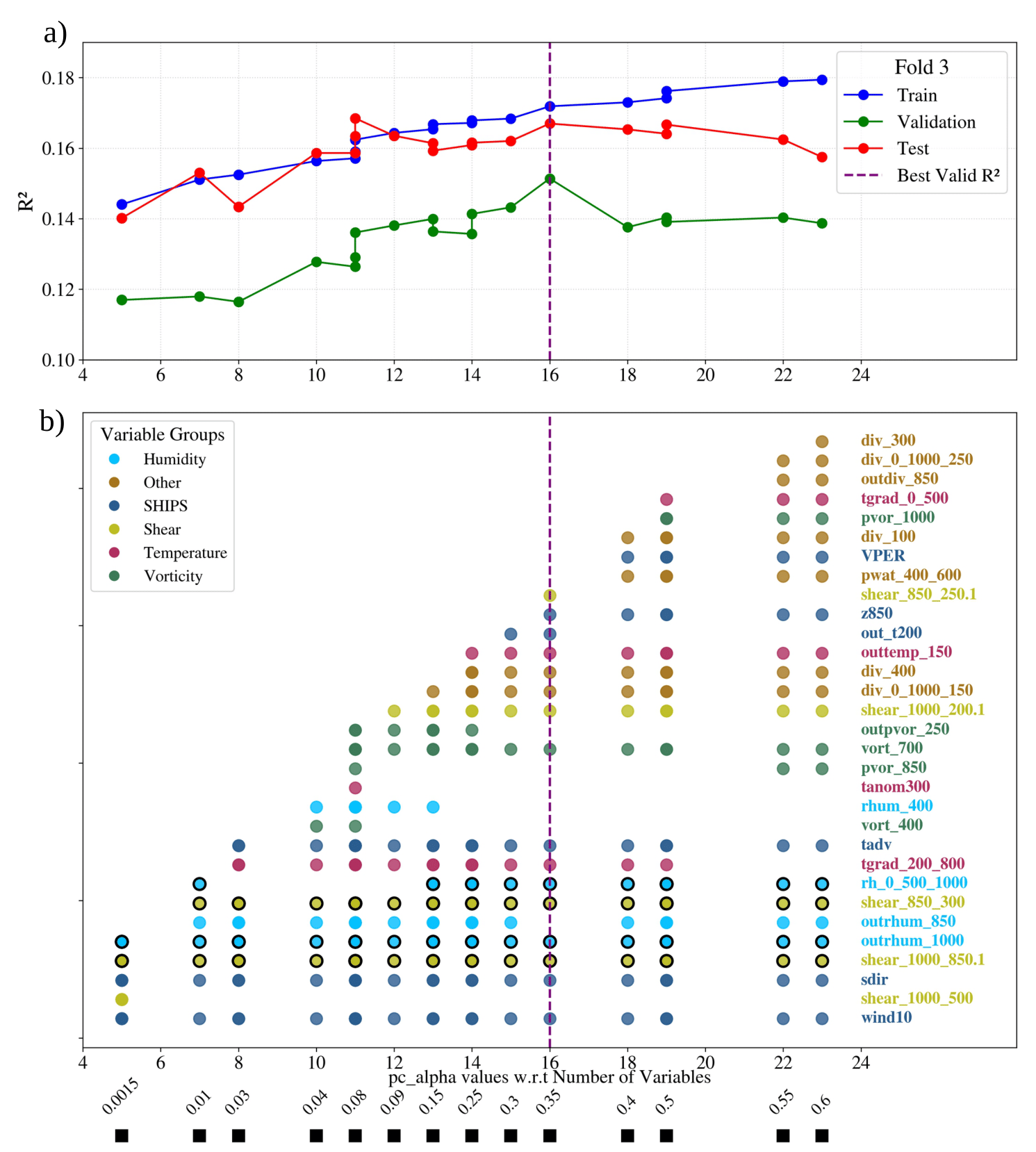}
    \caption{
Example results for the 24-hour intensity change forecast (DELV24) from Fold~3 using the SHIPS+ERA5 predictor set for SHIPS predictors. (a) Coefficient of determination $R^2$ on training, validation, and test sets plotted against the number of selected predictors, each point corresponding to a different value of the M-PC causal discovery hyperparameter \texttt{$pc_{\alpha}$} (bottom scale). The vertical dashed line indicates the configuration with the highest validation $R^2$. (b) Variable selection abacus: each dot shows the presence of a predictor across the \texttt{$pc_{\alpha}$} range in the test set. Variables are colored by group (e.g., Original SHIPS predictors, Shear, Humidity), vertical dashed line marks the best validation score, and encircled dots highlight the occurrence of new shortlisted predictors for SHIPS.}
    \label{FIG3}
\end{figure*}

We now present results from the ERA5-based replication experiments, focusing on how the M-PC framework operates and how the predictor selections translate into statistical forecast skill using MLR. 
Fig. ~\ref{FIG3} illustrates one realization of the ERA5-based causal discovery experiments for the 24-hour intensity change forecast (DELV24), shown for a representative cross-validation fold (Fold~3) under the noASSUMPS (“kitchen-sink”) configuration. The top panel shows training, validation, and test $R^2$ scores as a function of the number of input predictors in the MLR, controlled by the M-PC hyperparameter ${pc}_{\alpha}$. Higher values of ${pc}_{\alpha}$ correspond to a less strict statistical threshold for rejecting independence, allowing more predictors to be included in the causal model whereas a lower ${pc}_{\alpha}$ is more strict, resulting in a more stringent selection of predictors.

The bottom panel presents the variable selection abacus, highlighting predictor presence and groupings across the range of ${pc}_{\alpha}$ values. The encircled dots highlight the predictors included in the final shortlisted set, identified based on their consistent selection frequency across multiple folds.
The same procedure was repeated for seven cross-validation folds, all forecast lead times, and both experimental setups. Given the volume of results, Fig. ~\ref{FIG3} presents one representative fold for the noASSUMPS experiment, while the best performing folds for lead times of up to 120 hours are provided in the Supplementary Information (Figures~S1--S10). The best potential causal predictors that complement or improve upon the baseline SHIPS predictors for TC intensity change forecasting tend to be consistently shortlisted by the causal feature selection framework across different data splits and validated configurations, ensuring that the predictors perform well over a wide range of TC cases. 

\subsection{Causal feature selection outperforms baseline methods\label{subsec:Results_causalfeatureselectionoutperforming}}

\begin{figure*}[ht]
    \centering
    \includegraphics[width=0.8\textwidth]{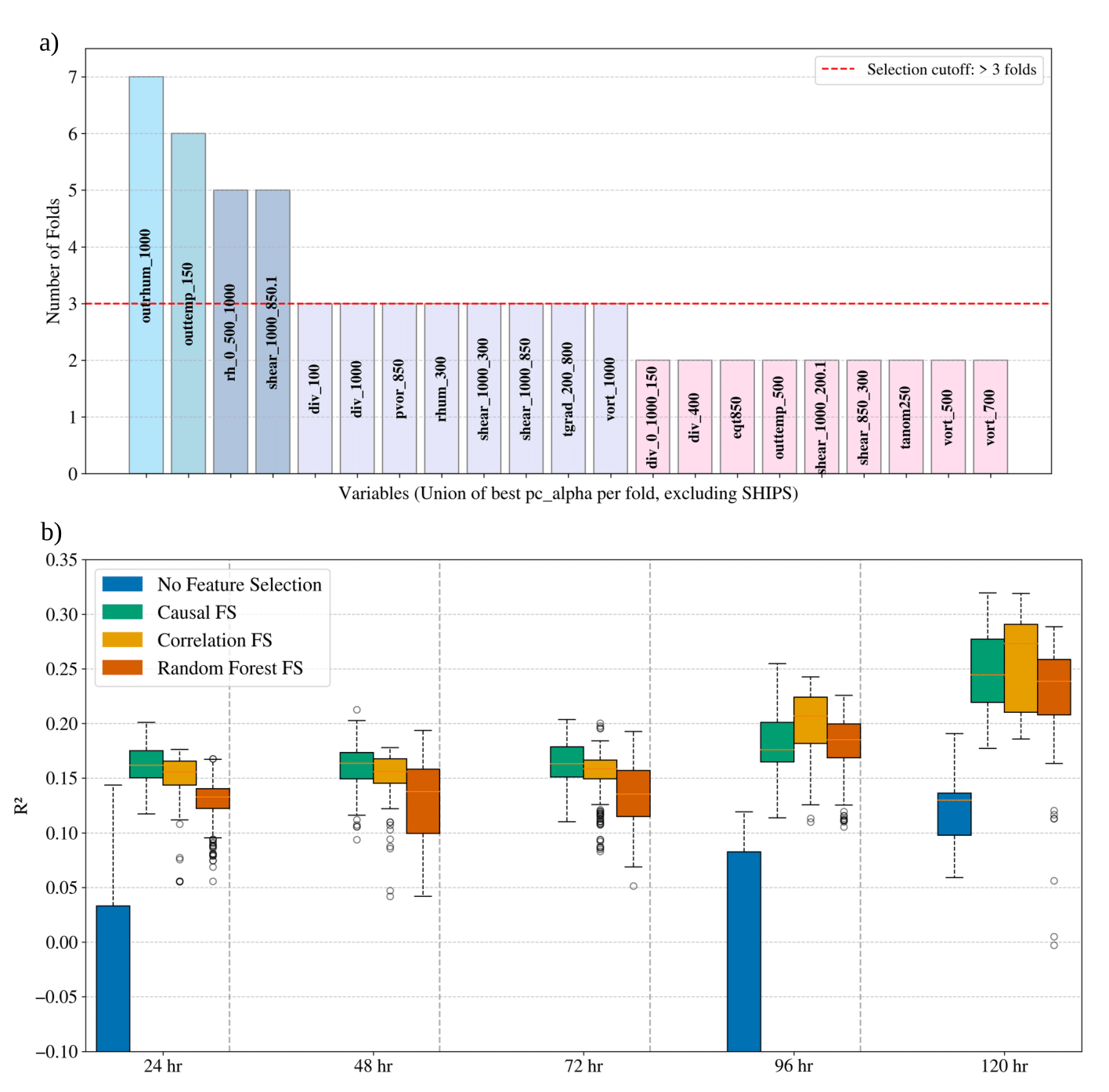}
    \caption{
    Summary of results for the 24-hour intensity change forecast (DELV24) using SHIPS+ERA5 predictors for the noASSUMPS experiment. (a) Bar plot showing the frequency of each variable’s selection across the best models from all seven cross-validation folds. A red dashed line marks the threshold (more than 3 folds) used to shortlist robust predictors for inclusion in the final SHIPS+ list. (b) Boxplot comparing test R² values for target DELV for each lead time (24, 48, 72, 96, 120 hours) for experiments with kitchen-sink approach (without link assumptions) across four feature selection strategies: causal discovery, correlation ranking, random forest importance, and no feature selection (all features). Causal feature selection yields the highest median R² until 72-hour lead time, showing improved generalization in a purely statistical prediction setup.}
    \label{fig:fig4}
\end{figure*}

Fig. ~\ref{fig:fig4} summarizes the process and outcomes of identifying robust predictors using our causal feature selection pipeline for the 24-hour intensity change forecast (DELV24).
Panel~(a) shows how often each candidate predictor was selected in the seven cross-validation folds. By applying a clear threshold, which requires a predictor to appear in more than three folds, we filter out variables with inconsistent contributions and focus on predictors with stable links to intensity change. 
Panel~(b) compares test $R^2$ scores for DELV across the four feature selection strategies at different lead times for the noASSUMPS experiment. Corresponding train and validation results are shown in Fig.~S14, while results for the withASSUMPS experiment (train, validation, and test) are provided in Fig.~S13 of the Supplementary Information. Causal feature selection achieves the highest median skill, outperforming correlation ranking, random forest importance, and the no feature selection baseline, demonstrating improved generalizability with causal predictors. Comparing the causal and correlation feature selection methods, we observe that causal yields fewer outliers in prediction errors compared to the correlation-based approach, yet displays slightly larger uncertainty bounds. This is because the PC-based causal discovery method tends to identify stable, interventionally relevant relationships rather than purely predictive ones. When applied to a system with underlying nonlinear dynamics, the PC-based model may underfit complex regions of the predictor space, leading to broader prediction intervals, which reflects the model's caution in extrapolating beyond what the causal structure supports. This caution mitigates large, spurious prediction errors and results in fewer outliers and more robust predictions on unseen test cases. The trade-off between model stability and uncertainty shown here highlights the value of causal methods over correlation in building reliable statistical models for high-impact forecasting applications.
By extending this comparison across all lead times (Fig S11, S12 in the SI) and evaluating the consistency of variable selection, we ultimately shortlisted six predictors to be added to SHIPS. These variables reliably contribute to improved forecast skill for short-range intensity change prediction, particularly up to 72 hours. Beyond this time frame, the benefit of causal feature selection diminishes, emphasizing that while our method provides meaningful gains in the short-range, longer lead times will likely require integrating statistical models with dynamical forecast guidance to capture additional sources of predictability.
\subsection{Recommending causally relevant predictors\label{subsec:Results_suggestingpredictors}}
In the previous section, we established that the causal predictors improve the generalizability of statistical TC intensity models for shorter lead times up to 72 hours, but not for longer lead times (Fig. \ref{fig:fig4}b).  
\begin{table}[t]
\caption{Recommended additional predictors to the operational SHIPS model.}
\label{tab:predictors}
\centering
\footnotesize
\renewcommand{\arraystretch}{1.2} 
\begin{tabular}{llccp{6cm}} 
\hline
Variable Group & Replication Code & Lead Time & SHIPS Code & Variable Description \\
\hline

\multirow{3}{*}{Shear} 
  & Shear\_1000\_850   & 24h, 48h, 96h    & SHL0 & Vertical shear 1000--850 hPa, area-averaged 200--800 km. \\
  & Shear\_850\_300    & 24h, 48h         & SHMD & Vertical shear 850--300 hPa, area-averaged 200--800 km. \\
  & Shear\_1000\_850.1 & 48h, 72h, 120h   & SHL1 & Vertical shear 1000--850 hPa, area-averaged 200--1000 km. \\

\hline
\multirow{2}{*}{Humidity} 
  & Outrhum\_1000      & 24h, 48h         & R000 & Relative Humidity at 1000 hPa, area-averaged 200--800 km. \\
  & RH\_0\_500\_1000   & 48h, 72h         & R001 & Relative Humidity at 1000 hPa, area-averaged 0--500 km. \\

\hline
Potential Vorticity 
  & Outpvor\_500 & 72h & PVOR & Potential Vorticity at 500 hPa, area-averaged 200--800 km. \\
\hline
\end{tabular}
\end{table}

Here, we use the coefficient of determination ($R^2$) as the primary performance metric, where positive values indicate that the model explains some fraction of the variance in TC intensity change, while negative values indicate that the model performs worse than simply predicting the mean. To increase the robustness of the predictor list and reduce the uncertainties arising from model hyperparameters and cross-validation strategies \cite{sweet2023cross}, the variable lists from the best MLR models for different folds are aggregated as a summary variable list. Of the different variables in the list, only those that are chosen in more than half of the cross-validation folds (at least 4 times) will be considered as candidate features to be included in the operational SHIPS for testing. Applying this shortlisting criterion and comparing frequently repeated variables across experiments withASSUMPS and noASSUMPS (Fig. \ref{fig:fig4} a, Fig. S11, Fig. S12) yields a final shortlist of six final predictors that mostly describe lower and middle tropospheric vertical wind shear (e.g. SHL0, SHL1, SHMD), surface and boundary layer moisture (R000, R001), and midtropospheric potential vorticity conditions (PVOR) in the outer area of TC. The complete list of additional predictors recommended is provided in Table~\ref{tab:predictors}. The most frequently selected causal predictors describe the kinematic and thermodynamic environment in the TC outer core, which likely reflects the critical implication of TC outer rainbands on TC structure and intensity \cite{wang2009outer,yu2021asymmetric}.

The physical relevance of these variables to TC intensity change is assessed prior to operational testing. Low-tropospheric wind shear, which usually promotes linear convective organization rather than circular organization \cite{robe2001effect,finocchio2016idealized}, is more negatively correlated with the intensity of Pacific typhoons that occurred in active typhoon seasons than commonly used deep-layer shear \cite{wang2015statistical}. \citep{fu2019effect} further conclude that low-level shear produces quasiperiodic oscillations in the intensity of the TC that are related to the variability in the moisture of the boundary layer induced by the TC rainbands. TC intensification is generally associated with relative humidity above the boundary layer \cite{wu2012relationship}, as reflected by the inclusion of RHMD in the original SHIPS predictors. In contrast, the 1000 hPa relative humidity in the outer environment (R000) is a variable not used in operational models but available in the extended predictor list. This selection can be physically explained as moisture within the boundary layer reflects the balance between surface flux moistening and dry air ventilation at the top of the boundary layer, through downdraft ventilation in sheared, dry environments \cite{alland2021Ventilation}, which can strongly modulate convective buoyancy and intensification rates \cite{riemer2010VWS_TCintensification, fischer2023tale}. Causal selection therefore reveals relevant predictors missed by traditional screening, improving short-range skills. The addition of overlooked variables such as surface relative humidity along with mid-tropospheric humidity and shear helps capture multiscale processes that drive intensity change, supporting the idea that boundary layer moisture could play a key role in regulating TC intensities \cite{wadler2021thermodynamic} and the onset of RI \cite{chen2021boundary}. Mid-level potential vorticity in the outer region of TCs reflects the dynamical environment of the storm and may capture interactions with environmental PV anomalies known to influence TC intensity \cite{molinari1998potential}, including synoptic-scale influences such as trough interactions \cite{fischer2019climatological}.

\subsection{Added value of the nonlinear SHIPS+ models \label{subsec:Results_SHIPSvsSHIPS+}}
The previous sections identify six additional predictors causally related to TC intensity change. These are evaluated within the operational SHIPS framework using predictors derived from GFS analysis and forecast fields. As a first step towards a fair evaluation of operational model skill, we compare models trained with the original SHIPS predictors and the enriched SHIPS+, which includes the six predictors. For consistency with the ERA5 experiments, only the initial analysis time data are used to derive both SHIPS developmental and SHIPS+ datasets. 

\begin{figure*}[ht]
    \centering
    \includegraphics[width=0.9\textwidth]{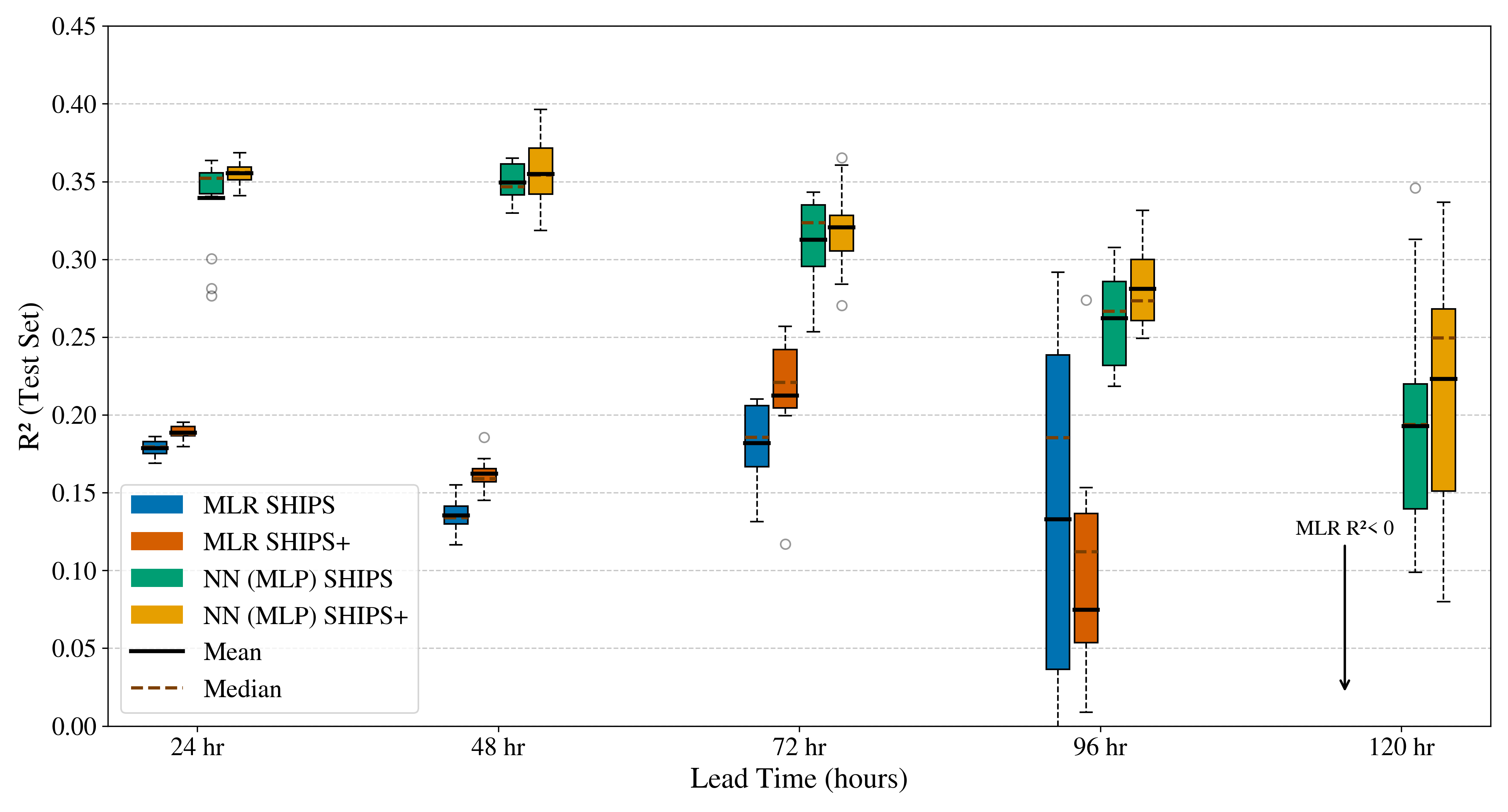}
    \caption{
Comparison of test $R^2$ values across forecast lead times (24–120 hours) for the original SHIPS predictors (blue/green boxes) and the expanded SHIPS+ predictors (orange/yellow boxes).
Both MLR and MLP runs are shown to illustrate the added value of nonlinear modeling.
Dashed brown lines indicate the median, and solid black lines mark the mean.
Overall, the MLP consistently outperforms the MLR, demonstrating improved skill when nonlinearity is captured, while the inclusion of additional predictors in SHIPS+ further enhances forecast performance, especially at shorter lead times. Note that the SHIPS+ MLR $R^2$ drops below 0 at the 120-hour lead time, which is indicated in the figure by a downward arrow.
}
    \label{fig:fig5}
\end{figure*}
Using $R^2$ as the performance metric, regression models trained with the additional causal predictors (yellow and orange boxes in Fig. ~\ref{fig:fig5}) outperform models trained with only the original SHIPS predictors (green and blue boxes in Fig. ~\ref{fig:fig5}) on unseen test TC cases, particularly at shorter lead times. This improvement is seen in both the linear MLR (blue and orange boxes) and nonlinear MLP (green and yellow boxes) models. However, the nonlinear SHIPS+ MLPs generally outperform their linear MLR counterparts. Since the SHIPS model is fully linear, we assess whether the superior performance of MLP models arises from their ability to capture nonlinear relationships between causal predictors and TC intensity change. Indeed, the lead time dependence of test skill improvements changes significantly in a nonlinear regression framework. The SHIPS+ MLRs show a clear lead time dependence with additional causal predictors, performing worse than the original SHIPS model beyond 72 hours, with negative $R^2$ at 120 hours. This lead time dependence is reduced in the nonlinear regression framework, where SHIPS+ outperforms SHIPS for all lead times. The results in Figure \ref{fig:fig5} imply that the relationship between the identified causal predictors and the rate of TC intensity change can be approximated linearly at shorter lead times but not at extended lead times, which makes MLRs incapable of utilizing additional information beyond 72 hours. Finally, the wider uncertainty ranges of the trained models at longer lead times suggest that these predictions cannot be adequately constrained with predictors derived solely from analysis time. This is a fundamental limitation of our predictive modeling setup, but it could be alleviated in operational settings by using dynamical model forecasts.

\begin{figure*}
    \centering
    \includegraphics[width=0.85\linewidth]{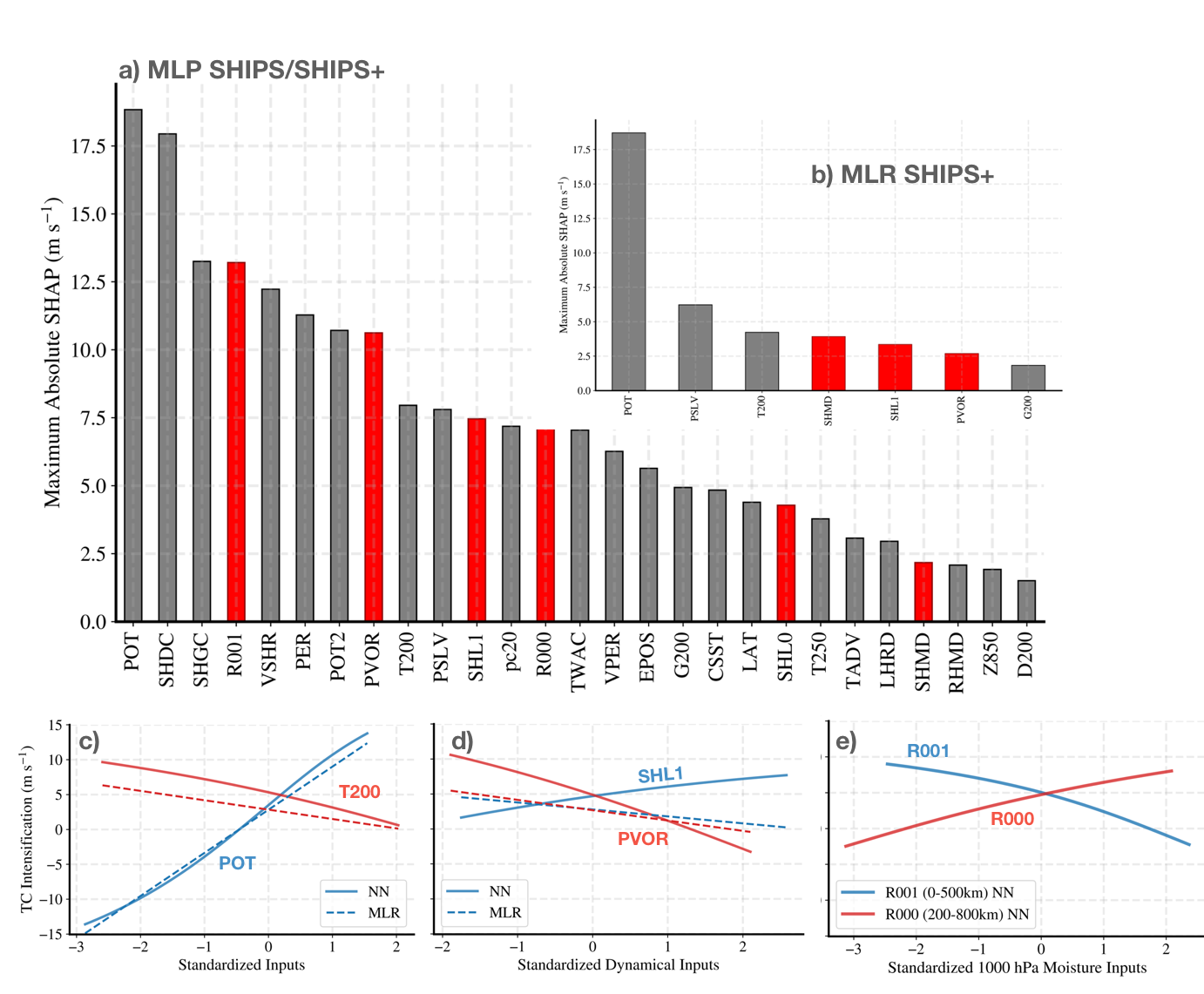}
    \caption{Predictor importance and dependencies for models trained on SHIPS+. (a–b) Global feature importance ranked by mean $\left|\textrm{SHAP}\right|$ for the MLP (a) and MLR (b). (c) SHAP dependence for baseline SHIPS predictors POT (potential intensity minus current intensity) and T200 (200 hPa temperature, 200–800 km-averaged). (d) SHAP dependence for causally selected predictors SHL1 (1000–850 hPa vertical wind shear, 200–1000 km-averaged) and PVOR (500 hPa potential vorticity, 200–800 km-averaged). (e) For near-surface humidity, the MLP learns opposite dependencies: negative with R001 (1000 hPa relative humidity, 0–500 km-averaged) and positive with R000 (1000 hPa relative humidity, 200–800 km-averaged). All predictors in panels c-e are standardized.}
    \label{fig:SHAP_analysis}
\end{figure*}

To better understand why the nonlinear SHIPS+ models outperform the other models, we conducted a SHAP (SHapley Additive exPlanations) analysis for the 24-hour lead time MLP and MLR models using both SHIPS and SHIPS+ datasets (see Fig. ~\ref{fig:SHAP_analysis}). SHAP values provide a model-agnostic interpretation of feature importance by quantifying the marginal contribution of each predictor to the model's output. To perform the SHAP analysis, we rely on Kernel SHAP \cite{kernel_shap_2017}, using 300 samples from the training dataset to retrieve the background signal and evaluating the 240 samples in the test set. Focusing on the differences between the linear MLR and the MLP models, we observe that of the six causal predictors, the MLP primarily utilizes the near-surface moisture predictors (red bars in Fig. \ref{fig:SHAP_analysis}a), while the SHIPS+ MLRs (Fig. \ref{fig:SHAP_analysis}b) use the causal shear predictors instead. A potential reason for MLPs and MLRs to select different causal predictors is that the relationship between near-surface moisture and short-term changes in TC intensity is too nonlinear for linear MLRs. The dependencies of the causal boundary layer moisture predictors (e.g., R001) on the MLR predictions are nonlinear (Fig.  \ref{fig:SHAP_analysis}e). Interestingly, the SHAP dependence plots reveal that the MLP learns opposing relationships for inner-core and outer-core near-surface humidity (Fig. ~\ref{fig:SHAP_analysis}e): increases in R001 (0–500 km) are associated with weaker predicted intensification, whereas increases in R000 (200–800 km) support stronger intensification. This result is unexpected, as higher inner-region moisture is generally associated with intensification. This contrast indicates that the MLP is leveraging radial gradients in boundary-layer moisture, a signal that the linear MLR framework does not capture. Apart from prioritizing additional nonlinear dependencies, the SHIPS+ MLPs could also overperform their MLR counterparts by learning different dependencies for the existing SHIPS predictors. Figure \ref{fig:SHAP_analysis}d presents two examples in which this is the case, where low-level wind shear (SHL1) and mid-tropospheric potential vorticity (PVOR) have a much stronger dependence on MLP predictions than the muted ones for MLR models. The negative correlation between intensification and PVOR might reflect the negative impact of TC outer rainbands on intensification \cite{wang2009outer,yu2021asymmetric}. Furthermore, MLPs learned a dependence between SHL1 and short-term intensity evolution that is strikingly different from the MLR dependence. Although intensification predictions in both the nonlinear and linear regression frameworks have a similar dependency on potential intensity (POT), which is the most critical predictor in both types of models, the difference in the selection of other leading predictors and their respective learned dependencies contribute to significant differences in the generalizability of the model to test TCs (Fig. ~\ref{fig:fig5}).

\subsection{Linear vs Nonlinear Models with Causal Predictors: Case Study\label{subsec:Results_casestudy}}
\begin{figure*}[t]
    \centering
    \includegraphics[width=0.95\textwidth]{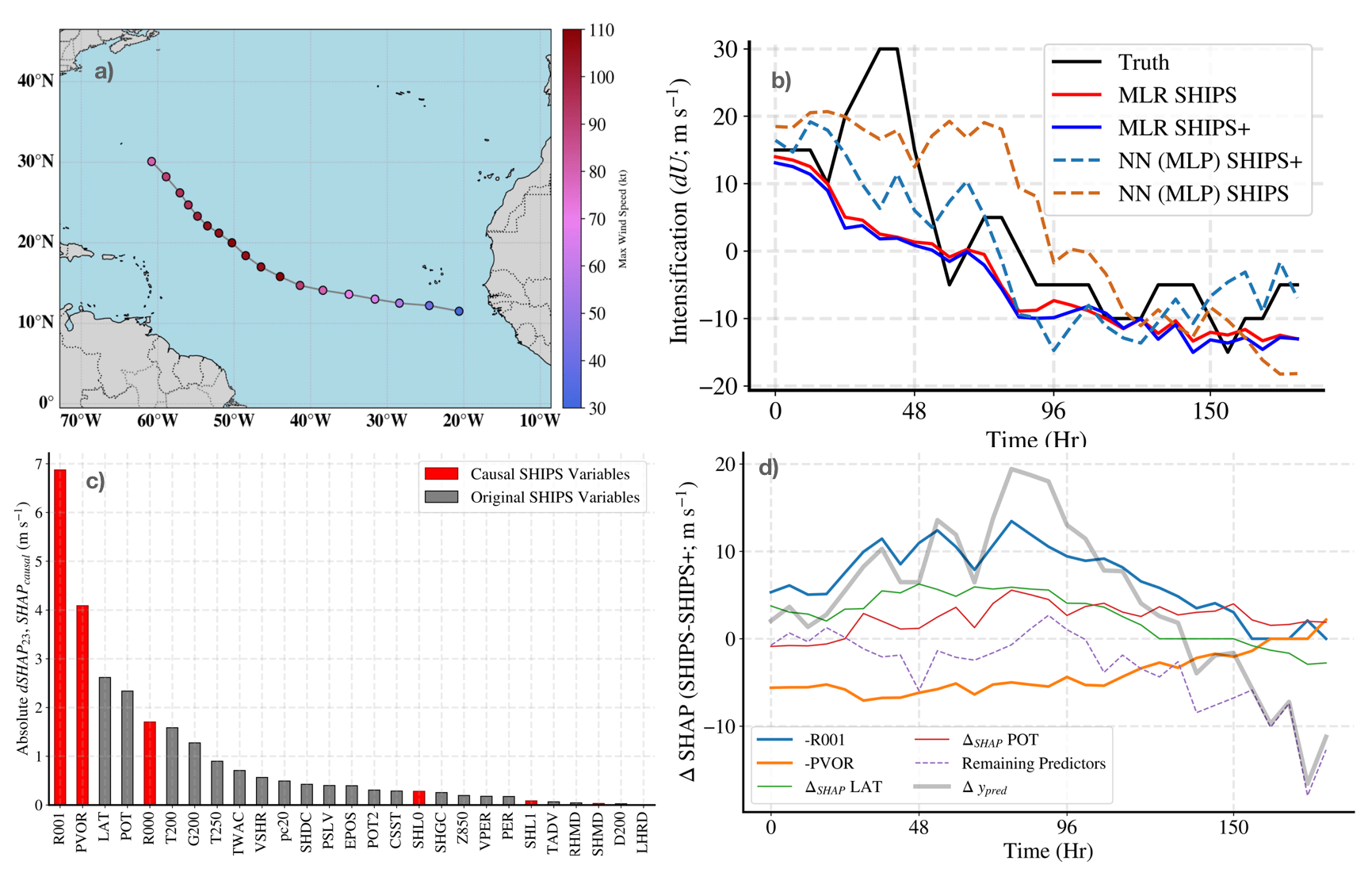}
    \caption{
Hurricane Larry (2021): (a) Best track from IBTrACS showing the time used for testing (b) performance comparison of MLR vs. MLP comparing IBTrACS 24-hour intensity change (truth) to predictions with (SHIPS+) and without causal predictors (SHIPS). Overall, the MLP consistently outperforms the MLR, with the addition of predictors improving the performance notable from 54 hours. (c) Lifetime-maximum absolute SHAP values highlighting the dominant role of the added predictors (R001, PVOR). (d) Time-series decomposition of $\Delta y_{\mathrm{pred}}$ showing R001 and PVOR as the main contributors, with smaller adjustments from LAT and POT.}
    \label{fig:case_study}
\end{figure*}

To illustrate the role of causally relevant predictors, we present Hurricane Larry (2021) as a case study. We selected Larry because, among the TCs in the independent test set, it showed the largest increase in predictive skill (R²) when using SHIPS+ MLPs with causal predictors compared to the operational SHIPS model. This allows us to investigate which predictors drive the improvement in forecast performance. Larry was a long-lived and intense Cape Verde hurricane that underwent a period of RI over the tropical Atlantic before eventually making landfall in Newfoundland as a Category 1 hurricane. For this analysis, we truncate the time series before its extratropical transition, focusing solely on the tropical phase. Fig. ~\ref{fig:case_study}  shows the best track of Larry during the evaluation period (a), and a comparison of model performance (b) between MLR and MLP, both with and without the addition of causally selected predictors. MLR SHIPS and MLR SHIPS+ both generally under-predict intensity change throughout the period, showing less variability and weaker response to fluctuations. Both versions of the Multi-Layer Perceptron (MLP) models with SHIPS and SHIPS+ (causally selected features) are consistently closer to the ground truth (black line) than the corresponding MLR models.
MLP SHIPS generally over-predicts TC Larry's intensification, except between 24 to 48 hours. Adding SHIPS+ dampens this overestimation, bringing it closer to the true signal, especially after 54 hours. In contrast, the performance gain from adding causal variables to MLR is minimal, indicating that linear models cannot fully leverage the additional causal features, highlighting the limitations of linear regression to capture TC dynamics, while the MLP models show a more dynamic response to intensity fluctuations.

To assess the role of causal predictors, we test whether the SHIPS$+$ MLP’s gains arise from learning new relations for the added predictors or simply from reweighting existing SHIPS predictors. Therefore, we apply the SHAP-based decomposition of \citep{grundner2021deep} (Appendix~B) to compare the SHIPS model $f$ and the SHIPS$+$ model $g$. By SHAP additivity, the prediction difference can be decomposed as follows:
\begin{equation}
\label{Eq:SHIPSdiff}
\begin{split}
\Delta y_{\mathrm{pred}}=B_f-B_g
&+\underbrace{\sum_{i=1}^{N_{\mathrm{SHIPS}}}\bigl(\mathrm{SHAP}_i^{f}-\mathrm{SHAP}_i^{g}\bigr)}_{\text{common predictors}}\\
&-\underbrace{\sum_{i=1}^{N_{\mathrm{causal}}}\mathrm{SHAP}_i^{g}}_{\text{added (causal) predictors}}\,
\end{split}
\end{equation}
where $B_f$ and $B_g$ are the SHAP ``background values'' (mean training-set predictions) for SHIPS and SHIPS$+$, respectively. The first sum aggregates the change in SHAP contributions for the $N_{\mathrm{SHIPS}}$ predictors present in both models; the second subtracts the SHAP contributions of the $N_{\mathrm{causal}}$ predictors included only in SHIPS$+$.

For Hurricane Larry, lifetime-mean absolute SHAP values (Fig. \ref{fig:case_study}c) indicate that Eq.\ref{Eq:SHIPSdiff} is well approximated by the contributions of two added predictors (R001 and PVOR) plus adjustments to two existing predictors (LAT and POT). The time-series decomposition (Fig. \ref{fig:case_study}d) confirms that R001 and PVOR are the two most important contributors to $\Delta y_{\mathrm{pred}}$, with smaller adjustments from LAT and POT and a small residual from all other features over the period of interest (30-120 hours). This lends credence to the interpretation that SHIPS$+$ improvements primarily arise from the added causal predictors. Consistent with the partial dependence for R001 (Fig. \ref{fig:SHAP_analysis}e), SHIPS underestimates the weakening of Larry after 54 hours because the original predictors under-represent the suppressing influence of inner-domain near-surface humidity; including R001 mitigates this bias.

\subsection{Testing the new predictors in operational SHIPS\label{subsec:Results_operationaltest}} 

As described by \cite{demaria2022national}, many new predictors have been added to SHIPS since the operational version was first implemented in 1991. As a preliminary test of how the research results presented above might improve the operational SHIPS model, the six potential new predictors listed in Table~\ref{tab:predictors} were evaluated using the standard SHIPS procedures for annual updates. For this test, the North Atlantic data from 1982-2021 were used for training and the 2022-2024 cases were used for testing. This procedure does not use validation data because the SHIPS prediction coefficients are uniquely determined from the MLR fit to the training data. SHIPS uses GFS model fields so the predictors in Table~\ref{tab:predictors} were recalculated from the GFS and added to the 28 predictors in the 2025 operational SHIPS model. 

As in SHIPS development, candidate variables are added to the developmental set and undergo a predictor-screening test for statistical significance with respect to intensity changes at 6-hour increments to lead times of 168~hours. Training follows the SHIPS perfect-prognosis protocol: predictors are averaged over each forecast interval and sampled from GFS \textit{analysis} fields about the future best-track positions at the valid times. In operations, the same predictors are computed from GFS \textit{forecast} fields along the National Hurricane Center (NHC) forecast track. Note that the causal predictor identification used in this manuscript (Table~\ref{tab:predictors}) differs by using only $t=0$ values about the current best-track position by design, to keep feature discovery in a purely statistical forecasting (not post-processing) setting and to make trse causal tests consistent with the regression task. We then apply the standard SHIPS screening and coefficient-update procedure described next. 
Three conditions are needed to pass the screening step: (1) When added one at a time, a new predictor must increase the variance explained by the model by at least 0.2\% averaged over 5 consecutive forecast intervals; (2) The regression coefficient must be significant at the 99\% level for at least 5 forecast intervals; (3) All predictors that pass steps 1 and 2 are added to the training sample and the coefficients are recalculated. The regression coefficients for each predictor must still be significant at the 99\% level for at least five forecast intervals. If a predictor does not pass step 3, the least significant predictor is removed and then step 3 is repeated until all retained predictors pass the significance test.  Results showed that SHMD, R001 and PVOR passed steps 1 and 2 for the 1982-2021 North Atlantic training sample. When all three were added, they also passed step 3. 
The next step in testing new SHIPS predictors is to perform retrospective forecasts on independent cases using only the input that is available in real time (forecast tracks and GFS forecast fields). To allow for a fair comparison, SHIPS was trained on the 1982-2021 sample with the original 28 operational predictors (baseline) and then with the addition of the three new predictors and both versions were run on the 2022-2024 independent cases. This sample included 794 cases with a 12-hour forecast, which decreased to 114 cases by 168 hours since many TCs dissipate by seven days. The mean absolute error (MAE) of the intensity forecasts for the baseline SHIPS and with the three new predictors were calculated using the intensity in the final NHC best track as ``truth'' following the standard NHC forecast procedure \cite{CangialosiReinhartMartinez2024}. The verification sample includes tropical and subtropical cyclones but excludes the extratropical and pre-genesis stages.
\begin{figure*}[ht]
    \centering
    \includegraphics[width=0.8\textwidth]{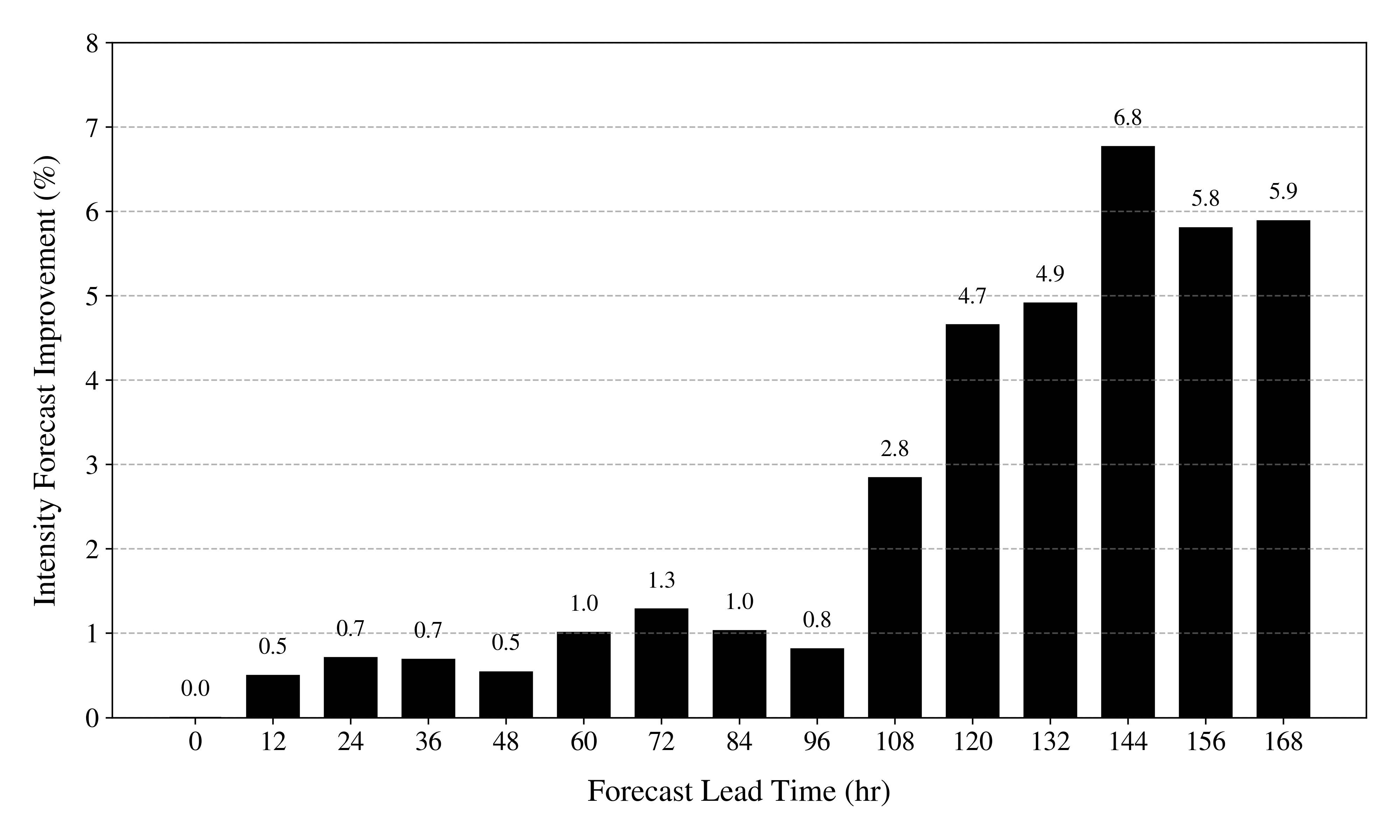}
    \caption{
    The improvement in the SHIPS intensity forecasts with the three new predictors relative to baseline for independent 2022--2024 Atlantic basin forecasts with real-time input.
    }
    \label{fig:Operational_SHIPS}
\end{figure*}

Fig. ~\ref{fig:Operational_SHIPS} shows the percent improvement (reduction in MAE) of SHIPS with the new predictors relative to baseline. SHIPS forecasts improved at all forecast times, with the largest improvements at longer forecast times. The relatively small improvements at short lead times reflect the dominant role of the persistence term in operational SHIPS, together with strong constraints imposed by initial conditions and directly assimilated observations, which limit the contribution of GFS-based environmental predictors. As forecast lead time increases, the influence of persistence and initial conditions weakens, while large-scale environmental controls become more important, allowing the newly identified predictors to produce larger gains in forecast skill. The improvements at 120-168 hours were statistically significant at the 90\% level using a standard statistical test that accounts for serial correlation. Although causal discovery used only $t=0$ values, the longer-lead time gains seen in Fig. ~\ref{fig:Operational_SHIPS} likely reflect the statistical–dynamical use of forecast fields in operations, motivating extension of the discovery procedure to predictors evaluated during the forecast period. The three new predictors (SHMD, R001 and PVOR) will be considered for implementation in future operational versions of SHIPS.   

\section{Conclusions}
To improve operational SHIPS, we introduced a multidata PC causal discovery framework to discover causally relevant predictors that drive North Atlantic TC intensity changes. First, we conducted tests by replicating the SHIPS predictors with the ERA5 reanalysis: we expanded the predictor list by testing combinations of dynamic and thermodynamic variables across different vertical layers and storm radial areas. M-PC tests using this augmented dataset revealed additional predictors that consistently demonstrated clear links to TC intensity change. Compared to other feature selection baselines, the causal approach performed best for short-range lead times (24–72 hours), highlighting its ability to isolate physically meaningful predictors while filtering out spurious relationships. Through this process, we shortlisted six key variables that were not currently used as operational SHIPS predictors. One of the new predictors, the surface level relative humidity for the outer area of the TC (R000), stands out because it is (1) already part of the extended SHIPS predictors list; and (2) has been linked to TC intensification and RI across multiple basins \cite{wu2012relationship,richardson2020evaluating,wang2020tropical}. The inclusion of other predictors, such as low-level shear and mid-tropospheric potential vorticity, underscores the importance of multiscale dynamical interactions and moisture–vorticity coupling processes that are often missed by traditional statistical frameworks.

This expanded predictors set, termed ``SHIPS+'', was rigorously tested by cross-validation across multiple folds and lead times. When using GFS analysis-time data from the SHIPS developmental dataset, incorporating these causal predictors consistently improved the forecast skill for short-range lead times up to 72 hours. The comparison of MLR and nonlinear MLP models further revealed that while linear models capture much of the intensity changes at shorter lead times, nonlinear interactions become increasingly important as lead time increases. The MLP consistently outperforms the MLR for all lead times and shows improved skill for SHIPS+ over the original SHIPS baseline, underscoring the value of combining causal feature selection with non-linear regression to better represent the evolving dynamics of TC intensification. This result suggests that the operational SHIPS model could be improved by replacing the MLR with a nonlinear method such as MLP in future versions.

The Hurricane Larry (2021) case study illustrated the added value of causally relevant predictors. Although both MLR and MLP models benefit from causal features, the nonlinear MLP better leverages these predictors to capture dynamic intensity fluctuations, including RI and decay phases, even though the predictors were selected using a \emph{linear} causal discovery procedure. In contrast, linear models tend to under-predict variability and respond less effectively to short-term changes, highlighting the advantage of nonlinear approaches in capturing complex TC behavior. These findings are consistent with the operational SHIPS tests, showing improvements at all lead times and the largest MAE reductions at 120–168 hours (Fig. \ref{fig:Operational_SHIPS}).

Although purely statistical models already benefit from causal feature selection and nonlinear methods, our results also point to their limitations at longer lead times, where static predictors become less informative. In addition, causal discovery and operational validation are constrained by the resolution and variable availability of ERA5 and GFS analyses, which may limit the representation of inner-core processes and exclude physically relevant predictors such as surface latent heat fluxes. However, in a statistical-dynamical framework such as operational SHIPS, where the GFS forecast model provides the evolving environmental fields, these limitations can diminish, as evidenced by the largest improvements at 120–168~hours. A promising next step is to derive causally selected predictors directly from dynamical forecast output and along forecast trajectories. This would allow their seamless integration into statistical-dynamical frameworks such as SHIPS, potentially improving the forecast skill for RI and addressing one of the main sources of uncertainty in operational TC intensity prediction.

\section*{Acknowledgments}
The authors thank Dr. Andreas Gerhardus (DLR, Germany) for the initial guidance on causal experiments with replicated data. The authors also acknowledge the DCSR at UNIL for providing computational resources and technical support. We thank the two anonymous reviewers for their detailed reviews and constructive suggestions. Tom Beucler acknowledges support from the Swiss National Science Foundation (SNSF) under Grant No. 10001754 (``RobustSR'' project). 

\paragraph{Author Contributions}
Conceptualization: S.G.S., T.B., F.I.T., J.R., M.D.;  
Methodology: J.R., T.B., S.G.S., F.I.T., M.S.G., M.D., K.M.;  
Data curation: S.G.S., F.I.T., M.S.G., T.B., M.M., M.D.;  
Data visualization: S.G.S., F.I.T., T.B., M.D.;  
Writing—original draft: S.G.S., F.I.T.;  
All authors contributed to writing and review. All authors approved the revised manuscript.

\paragraph{Competing interest}
The authors declare that no competing interests exist. 

\section*{Data availability}
The causal discovery package and tutorials are freely available in the \href{https://github.com/jakobrunge/tigramite}{Tigramite} GitHub repository. All the scripts, tutorials to replicate the experiments and preprocessed data for this study are available at \href{https://github.com/saranya8989/2025_CausalSHIPS}{Causal-SHIPS} and have been archived in \href{https://doi.org/10.5281/zenodo.17242382}{Zenodo}. Large datasets not included in the GitHub repository for the Part 1 and Part 3 scripts are available at https://doi.org/10.5281/zenodo.17241222. The North Atlantic TC data are from the \href{https://www.ncei.noaa.gov/products/international-best-track-archive}{IBtrACS} data archive. ERA5 datasets, including multiple and single pressure levels, were obtained from the Copernicus Climate Data Store at \href{https://cds.climate.copernicus.eu/}{cds.climate.copernicus.eu}. TC-PRIMED Data and documentation are available at \href{https://rammb-data.cira.colostate.edu/tcprimed/products.html}{this link}.  The SHIPS developmental dataset can be accessed at \href{https://rammb2.cira.colostate.edu/research/tropical-cyclones/ships/development_data/}{NESDIS SHIPS Developmental Data}.  All datasets are publicly available and can be freely accessed and reused under their respective data policies.

 \paragraph{Ethics Statement}
 The research meets all ethical guidelines.
 
\paragraph{Funding Statement}
This research was supported by the canton of Vaud in Switzerland. Tom Beucler acknowledges support from the Swiss National Science Foundation (SNSF) under Grant No. 10001754 (``RobustSR'' project).

\paragraph{Provenance}
We confirm that this study is original research, developed entirely by the authors, and has been submitted to Weather and Forecasting (AMS).

\paragraph{Supplementary Information}
Supplementary materials, organized into four sections, are provided with this manuscript.

\bibliographystyle{unsrt}  
\bibliography{main}  
\clearpage

\pagestyle{fancy}
\thispagestyle{empty}
\rhead{ \textit{ }}

\fancyhead[LO]{Causal Discovery to improve SHIPS}

\title{Supplementary Material: Multidata Causal Discovery for Statistical Hurricane Intensity Forecasting
}

\author{
  Saranya Ganesh S., Frederick Iat-Hin Tam, Milton S. Gomez, Tom Beucler \\
  Faculty of Geosciences and Environment,\\
  Expertise Center for Climate Extremes,\\ 
  University of Lausanne \\
  Lausanne, Vaud, Switzerland\\
   \And
  Marie McGraw, Mark DeMaria, Kate Musgrave \\
  Cooperative Institute for Research in the Atmosphere, \\ 
  Colorado State University, \\
  Fort Collins, Colorado\\ 
   \And
  Jakob Runge \\ 
  Department of Computer Science,\\
  University of Potsdam \\
  Potsdam, Germany \\
}

\maketitle

\setcounter{page}{1}

This Supplementary Information (SI) provides additional details and results supporting the main manuscript.

\begin{itemize}
   \item The full set of variables used as predictors in our tropical cyclone intensity prediction experiments, including the core SHIPS predictors, is provided in Tables~S1–S5: Table~S1 lists the original SHIPS predictors, Table~S2 their replication from ERA5, Table~S3 the additional inner-core variables (0–2$^{\circ}$), Table~S4 the outer-core variables (200–800 km, up to 1000 km), and Table~S5 the extended set of predictors from the TC PRIMED dataset. The SHIPS replication experiment uses a subset of predictors consistent with those available from ERA5 reanalysis and the TC PRIMED dataset, excluding variables such as PC20, PSLV, SST, and SHGC, except when directly compared with operational SHIPS runs.

   \item The causal discovery experiments were conducted using 24 statistical significance thresholds:
    \begin{align*}
    pc_{\alpha} \in \{&
    0.0001,\; 0.00015,\; 0.001,\; 0.0015,\; 0.01,\; 0.02,\; 0.03,\; 0.04,\\
    &0.05,\; 0.06,\; 0.07,\; 0.08,\; 0.09,\; 0.1,\; 0.15,\; 0.2,\\
    &0.25,\; 0.3,\; 0.35,\; 0.4,\; 0.45,\; 0.5,\; 0.55,\; 0.6 \}.
    \end{align*}

    \item Figures S1–S10, which are similar in format to Figure 3 in the main text, showthe fold with highest performance on the validation set (based on R²) out of the seven experiments, with and without link assumptions, across the five lead times (24, 48, 72, 96, and 120 hours).

    \item Figures S11–S12, which are similar in format to Figure 4a in the main text, show the frequency of each variable in the best models of all seven cross-validation folds for the experiments with and without SHIPS link assumptions for the target DELV at lead times from 24 to 120 hours. A cutoff of 3 is applied, where predictors are shortlisted if they appear at least 4 times across the 7 cross-validation experiments.

    \item Figures S13–S14, which are similar in format to Figure 4b in the main text, show box plots of coefficient of determination (R²) values comparing different feature selection methods for Train (top), Validation (middle), and Test (bottom) for DELV at lead times from 24 to 120 hours, with and without SHIPS link assumptions.

    \item Figure S15, which is similar in format to Figure 5 in the main text, show box plots of coefficient of determination (R²) values comparing different feature selection methods for the SHIPS+ dataset for Training (top) and Validation (bottom) sets at lead times from 24 to 120 hours, with and without SHIPS link assumptions.
\end{itemize}
\clearpage
\setcounter{table}{0}
\setcounter{figure}{0}
\renewcommand{\thetable}{S\arabic{table}}

\begin{table}[t]
\caption{SHIPS Developmental Dataset Variables}
\label{table:ships_dataset}
\centering
\footnotesize
\renewcommand{\arraystretch}{1.2}
\begin{tabular}{ll}
\toprule
\textbf{Variable} & \textbf{Description} \\
\midrule
DELV24/48/72/96/120 & Target: Max surface wind difference over 24/48/72/96/120 h \\
PMIN & Minimum central pressure of the system (hPa) \\
VMAX & Maximum wind speed (kt) \\
PER & Intensity change (kt) over prior 12 h \\
VPER & PER multiplied by VMAX \\
PC20 & Percent of GOES IR pixels colder than --20°C, area-averaged 50--200 km \\
SPDX & X-component of storm translational speed (kt) at forecast time \\
PSLV & Steering layer pressure center of mass (hPa) at forecast time \\
SST & Sea surface temperature at storm center (°C), time-averaged 0--48 h \\
POT & Potential intensity minus current intensity (kt), time-averaged 0--48 h \\
SHDC & 850--200 hPa vertical wind shear magnitude (kt), time-averaged 0--48 h \\
T200 & 200 hPa temperature (°C), 200--800 km area-avg, time-avg 0--48 h \\
T250 & 250 hPa temperature (°C), background-subtracted \\
EPOS & Parcel instability parameter from equivalent potential temperature (°C) \\
RHMD & Relative humidity (\%) in 500--700 hPa, 200--800 km area-avg \\
TWAT & Time tendency of average tangential wind within 500 km \\
Z850 & Relative vorticity ($10^{-7}$ s$^{-1}$), 0--1000 km area-avg \\
D200 & 200 hPa divergence ($10^{-7}$ s$^{-1}$), 0--1000 km area-avg \\
LHRD & SHDC multiplied by sine of latitude \\
VSHR & VMAX multiplied by SHDC \\
POT2 & Square of POT \\
SHGC & Generalized shear (kt) from 100--1000 hPa levels \\
SDIR & Deviation of 850--200 hPa shear direction from optimal (°) \\
TADV & Temperature advection between 850--700 hPa, area-avg 0--500 km \\
G200 & Temperature perturbation (°C) at 200 hPa, 200--800 km, time-avg 0--48 h \\
LAT & Latitude of TC center \\
\bottomrule
\end{tabular}
\end{table}

\begin{table}[t]
\caption{ERA5 Replication of SHIPS Predictors}
\label{table:era5_ships_predictors}
\centering
\footnotesize
\renewcommand{\arraystretch}{1.2}
\begin{tabular}{ll}
\toprule
\textbf{Variable} & \textbf{Description} \\
\midrule
DELV & Intensity change over 24/48/72/96 h \\
PMIN & Minimum central pressure (hPa) \\
VMAX & Max wind speed from ERA5 10m winds (kt) \\
PER & Intensity change (kt) over prior 12 h \\
VPER & PER multiplied by VMAX \\
SPDX & ERA5 x-component of storm translational speed (kt) \\
PSLV & ERA5 steering layer pressure center (hPa) \\
SST & ERA5 sea surface temperature (°C) \\
POT & ERA5 potential intensity minus current (kt) \\
SHDC & ERA5 850--200 hPa vertical shear (kt) \\
T200 & ERA5 200 hPa temperature (°C) \\
T250 & ERA5 250 hPa temperature (°C) \\
EPOS & ERA5 parcel instability from equivalent potential temperature (°C) \\
RHMD & ERA5 relative humidity (\%) in 500--700 hPa \\
TWAT & ERA5 time tendency of average tangential wind \\
Z850 & ERA5 850 hPa relative vorticity ($10^{-7}$ s$^{-1}$) \\
D200 & ERA5 200 hPa divergence ($10^{-7}$ s$^{-1}$) \\
LHRD & ERA5 shear magnitude (kt × 10) with vortex removed \\
VSHR & ERA5 VMAX × SHDC (kt$^2$) \\
POT2 & ERA5 POT squared (kt$^2$) \\
SHGC & ERA5 generalized shear (kt) \\
SDIR & ERA5 shear direction deviation (°) \\
TADV & ERA5 temperature advection (°C) \\
\bottomrule
\end{tabular}
\end{table}

\begin{table}[t]
\caption{Inner Core (0--200 km area-averaged) Variables from ERA5}
\label{table:inner_core_variables}
\centering
\footnotesize
\renewcommand{\arraystretch}{1.2}
\begin{tabular}{lll}
\toprule
\textbf{Variable} & \textbf{Description} & \textbf{Pressure Levels (hPa)} \\
\midrule
div & Horizontal divergence (s$^{-1}$) & 100, 200, 250, 300, 400, 500, 700, 850, 1000 \\
eqt & Equivalent potential temperature (K) & 1000, 200, 250, 300, 400, 500, 700, 850 \\
vort & Relative vorticity (s$^{-1}$) & 100, 150, 200, 250, 300, 400, 500, 700, 850, 1000 \\
pvor & Potential vorticity (PVU) & 100, 150, 200, 250, 300, 400, 500, 700, 850, 1000 \\
rhum & Relative humidity (\%) & 100, 150, 200, 250, 300, 400, 500, 700, 850, 1000 \\
gpot & Geopotential height (m) & 100, 150, 200, 250, 300, 400, 500, 700, 850, 1000 \\
temp & Air temperature (K) & 100, 150, 200, 250, 300, 400, 500, 700, 850, 1000 \\
\bottomrule
\end{tabular}
\end{table}

\begin{table}[t]
\caption{Outer Area (200--800 km area-averaged) Variables from ERA5}
\label{table:outer_area_variables}
\centering
\footnotesize
\renewcommand{\arraystretch}{1.2}
\begin{tabular}{lll}
\toprule
\textbf{Variable} & \textbf{Description} & \textbf{Pressure Levels (hPa)} \\
\midrule
outdiv & Horizontal divergence (s$^{-1}$) & 100, 200, 250, 300, 400, 500, 700, 850, 1000 \\
outeqt & Equivalent potential temperature (K) & 1000, 200, 250, 300, 400, 500, 700, 850 \\
outvort & Relative vorticity (s$^{-1}$) & 100, 150, 200, 250, 300, 400, 500, 700, 850, 1000 \\
outpvor & Potential vorticity (PVU) & 100, 150, 200, 250, 300, 400, 500, 700, 850, 1000 \\
outrhum & Relative humidity (\%) & 100, 150, 200, 250, 300, 400, 500, 700, 850, 1000 \\
outgpot & Geopotential height (m) & 100, 150, 200, 250, 300, 400, 500, 700, 850, 1000 \\
outtemp & Air temperature (K) & 100, 150, 200, 250, 300, 400, 500, 700, 850, 1000 \\
\bottomrule
\end{tabular}
\end{table}

\begin{table}[t]
\caption{TC PRIMED Variables used in this study.}\label{table:tcprimed_variables}
\centering
\footnotesize
\renewcommand{\arraystretch}{1.2} 
\begin{tabular}{lp{4cm}p{1.5cm}p{2cm}p{3.5cm}}
\hline\hline
\textbf{Variable} & \textbf{Description} & \textbf{Units} & \textbf{Radius (km)} & \textbf{Pressure Levels (hPa)} \\
\hline
shear / shear.1 & Vertical wind shear & m s$^{-1}$ & 200--800, 200--1000 & 850--200, 1000--300, 1000--500, 1000--700, 1000--850, 850--250, 850--300, 850--500 \\
tgrad & Temperature gradient & K km$^{-1}$ & 0--500, 200--800 & -- \\
pwat & Precipitable water & mm & 0--200, 200--400, 400--600, 600--800, 800--1000 & -- \\
div & Divergence & s$^{-1}$ & 0--1000 & 100, 150, 200, 250, 300, 400, 500, 700, 850, 1000 \\
vort & Vorticity & s$^{-1}$ & 0--1000 & 100, 150, 200, 250, 300, 400, 500, 700, 850, 1000 \\
geop & Geopotential height & m & 0--1000 & 100, 150, 200, 250, 300, 400, 500, 700, 850, 1000 \\
rh & Relative humidity & \% & 0--500 & 100, 150, 200, 250, 300, 400, 500, 700, 850, 1000 \\
tanom & Warm-core temperature anomaly & K & 0--15 km to 1500 km & 100, 150, 200, 250, 300, 400, 500, 700, 850, 1000 \\
\hline
\end{tabular}
\end{table}

\renewcommand{\thefigure}{S\arabic{figure}}

\begin{figure*}[h]
    \centering
    \includegraphics[width=\textwidth]{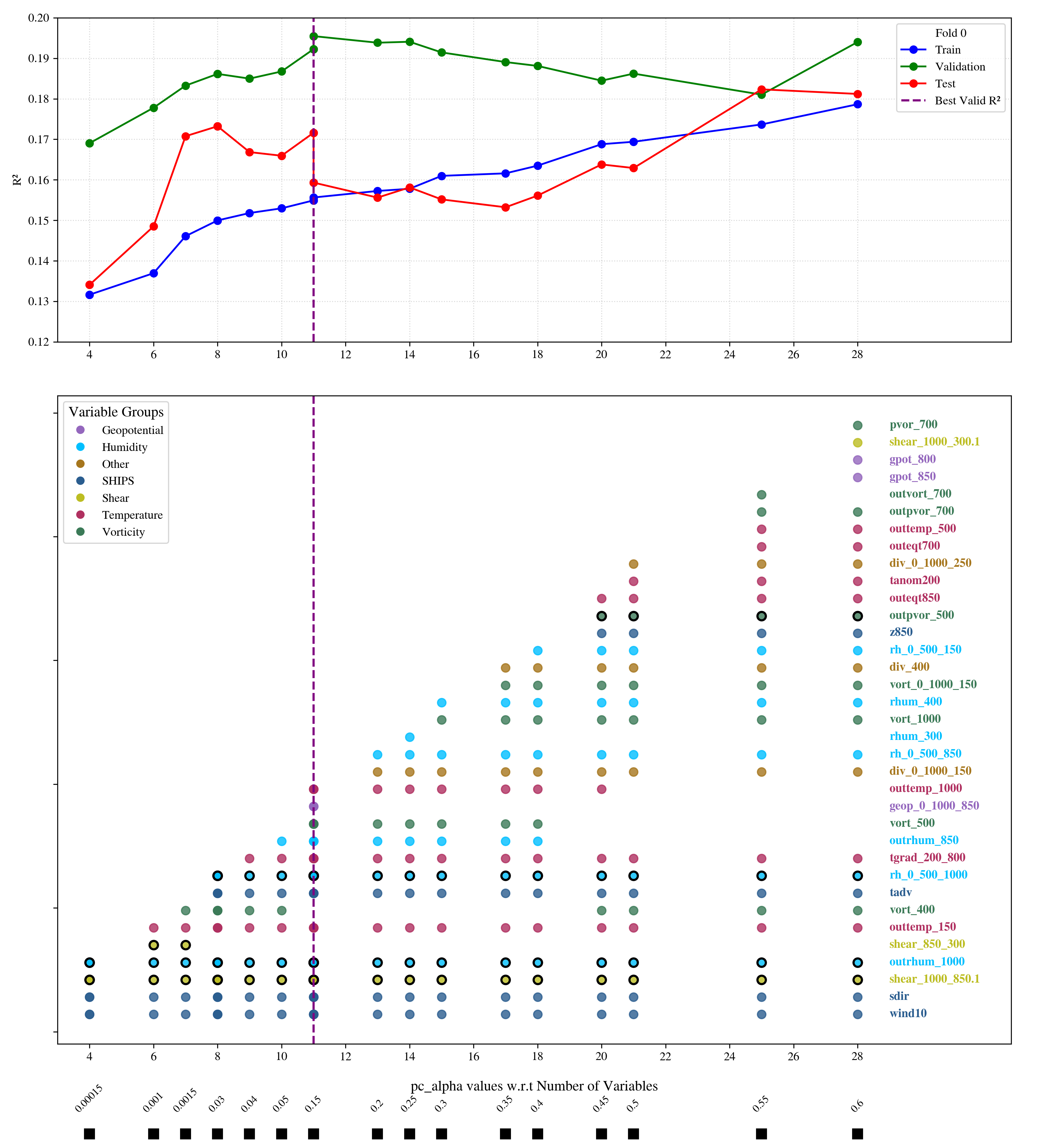}
    \caption{
Results for the 24-hour intensity change forecast (DELV24) from the best fold using the SHIPS+ERA5 predictor set \emph{without} SHIPS link assumptions. 
\textbf{Top panel:} $R^2$ scores on training, validation, and test sets plotted against the number of selected variables, each point corresponding to a different value of the M-PC1 causal discovery hyperparameter \texttt{pc\_alpha} (bottom scale). The vertical dashed line indicates the configuration with the highest validation $R^2$. 
\textbf{Bottom panel:} Variables selection: each dot shows the presence of a predictor across the \texttt{pc\_alpha} range. Variables are colored by group (e.g., SHIPS, Shear, Humidity).
}
    \label{FIGS1}
\end{figure*}

\begin{figure*}[h]
    \centering
    \includegraphics[width=\textwidth]{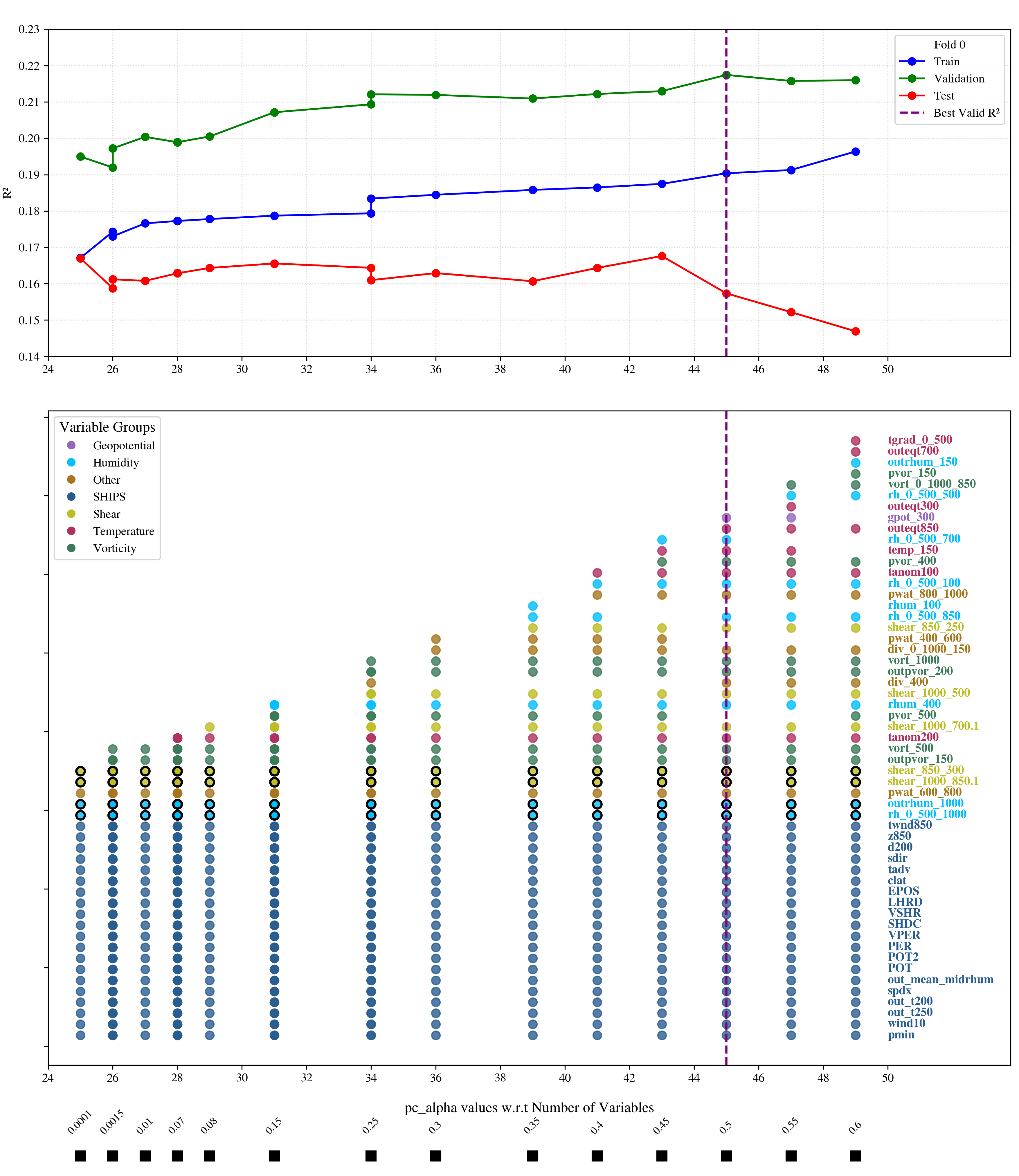}
    \caption{
Results for the 24-hour intensity change forecast (
DELV24) from the best fold using the SHIPS+ERA5 predictor set \emph{with} SHIPS link assumptions. Same as the Figure S1.
}
    \label{FIGS2}
\end{figure*}

\begin{figure*}[h]
    \centering
    \includegraphics[width=\textwidth]{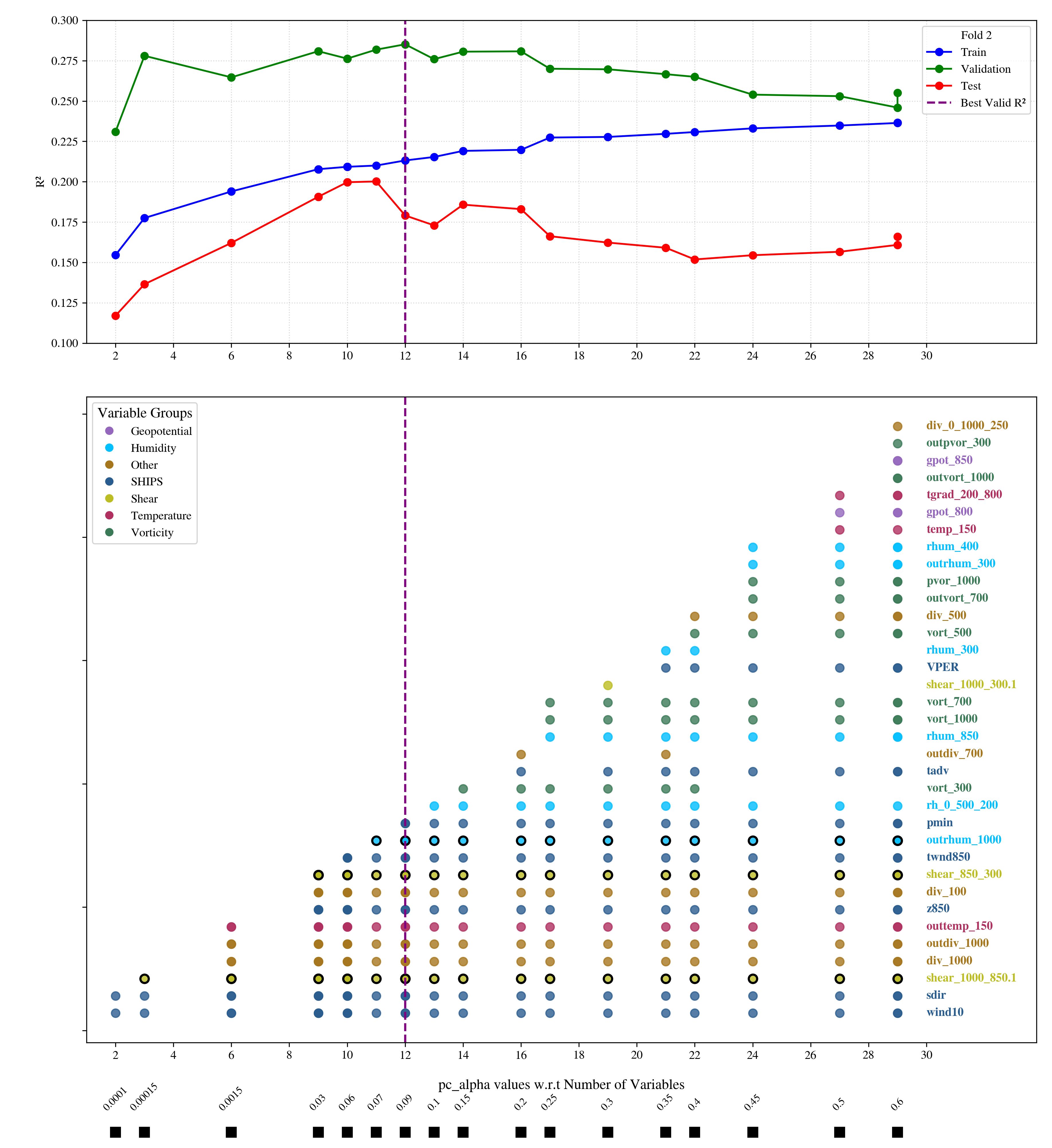}
    \caption{
Results for the 48-hour intensity change forecast (DELV48) from the best fold using the SHIPS+ERA5 predictor set \emph{without} SHIPS link assumptions. Same as the Figure S1.
}
    \label{FIGS3}
\end{figure*}

\begin{figure*}[h]
    \centering
    \includegraphics[width=\textwidth]{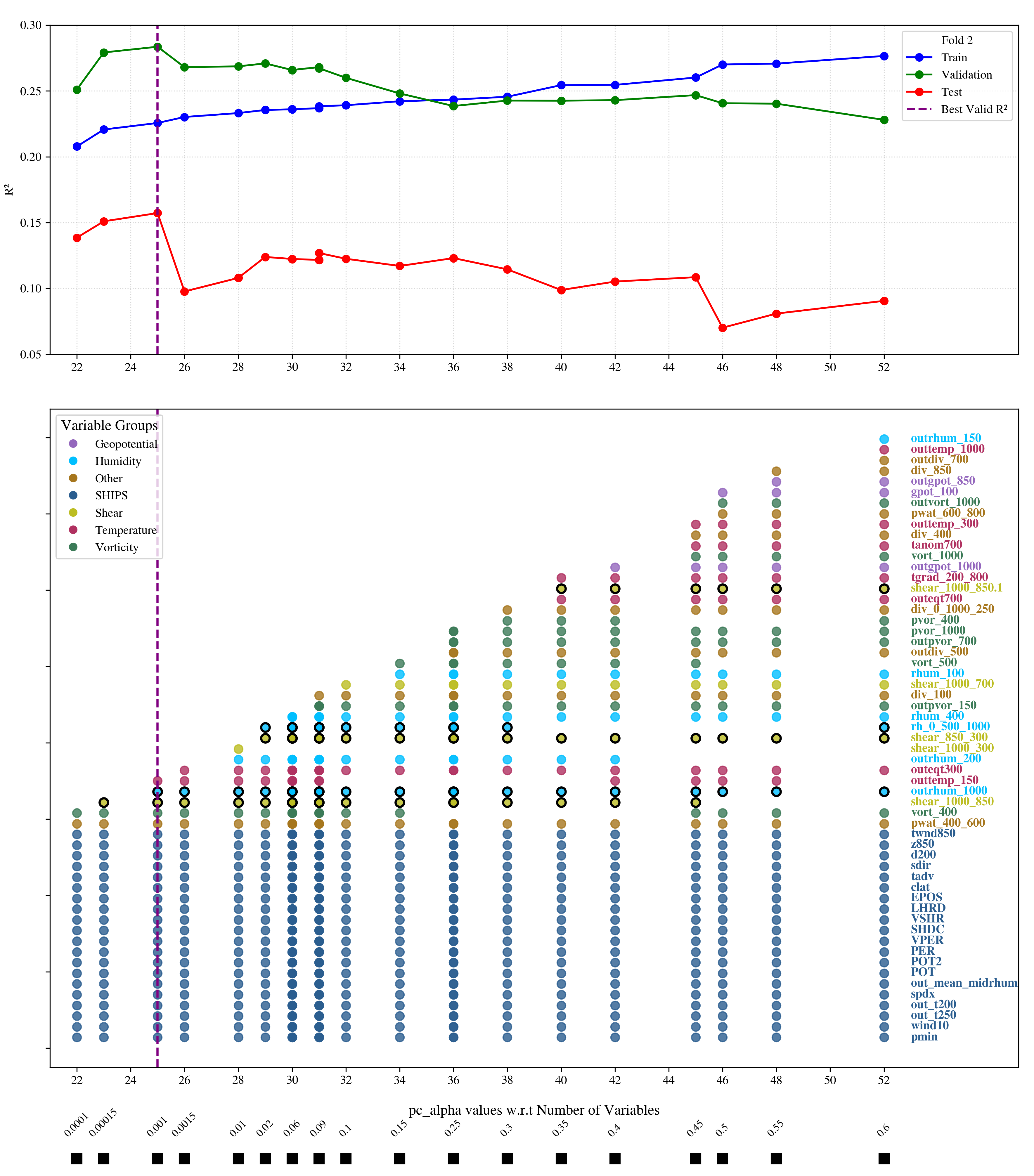}
    \caption{
Results for the 48-hour intensity change forecast (DELV48) from the best fold using the SHIPS+ERA5 predictor set \emph{with} SHIPS link assumptions. Same as the Figure S2.
}
    \label{FIGS4}
\end{figure*}

\begin{figure*}[h]
    \centering
    \includegraphics[width=\textwidth]{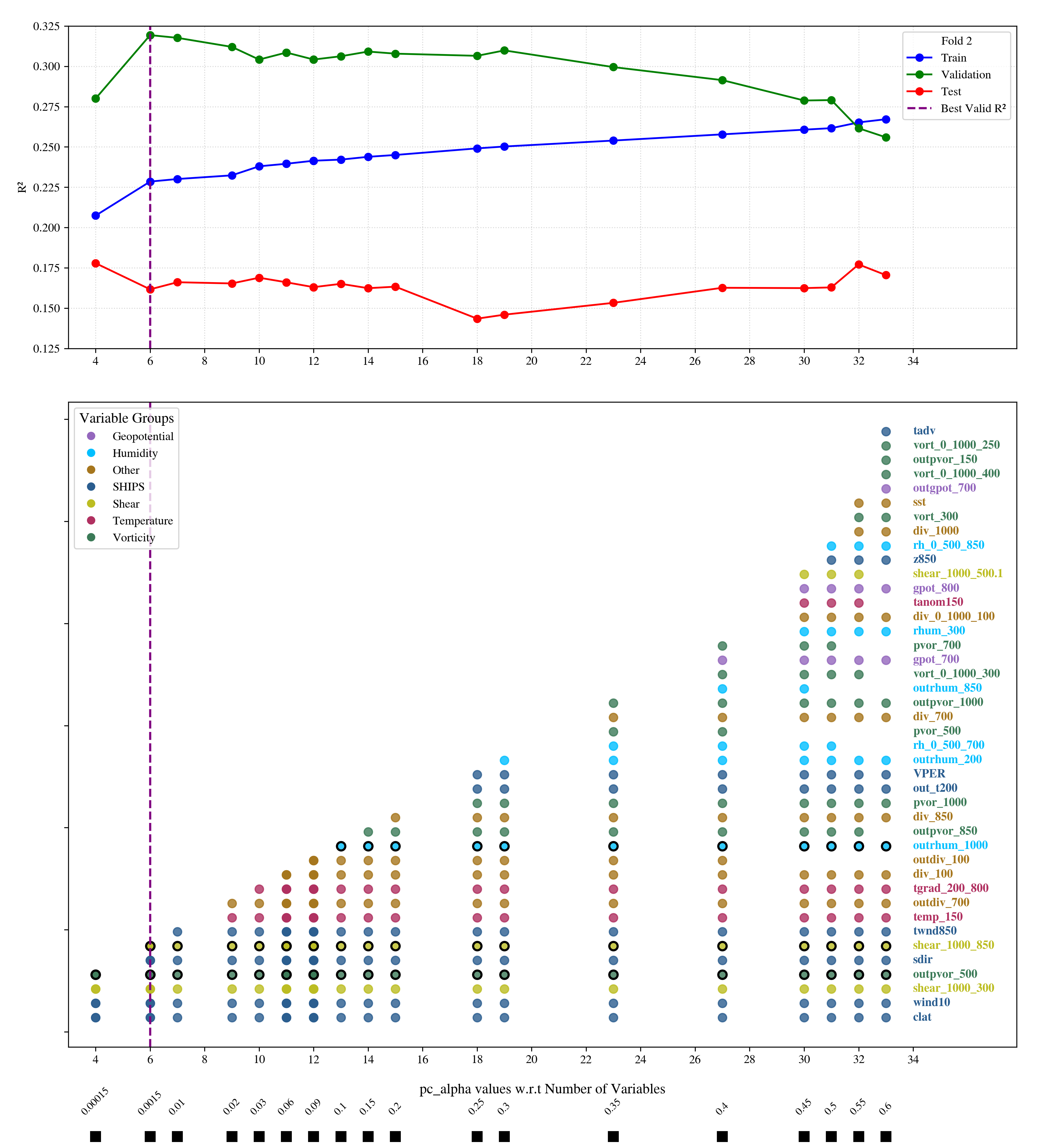}
    \caption{
Results for the 72-hour intensity change forecast (DELV72) from the best fold using the SHIPS+ERA5 predictor set \emph{without} SHIPS link assumptions. Same as the Figure S1.
}
    \label{FIGS5}
\end{figure*}

\begin{figure*}[h]
    \centering
    \includegraphics[width=\textwidth]{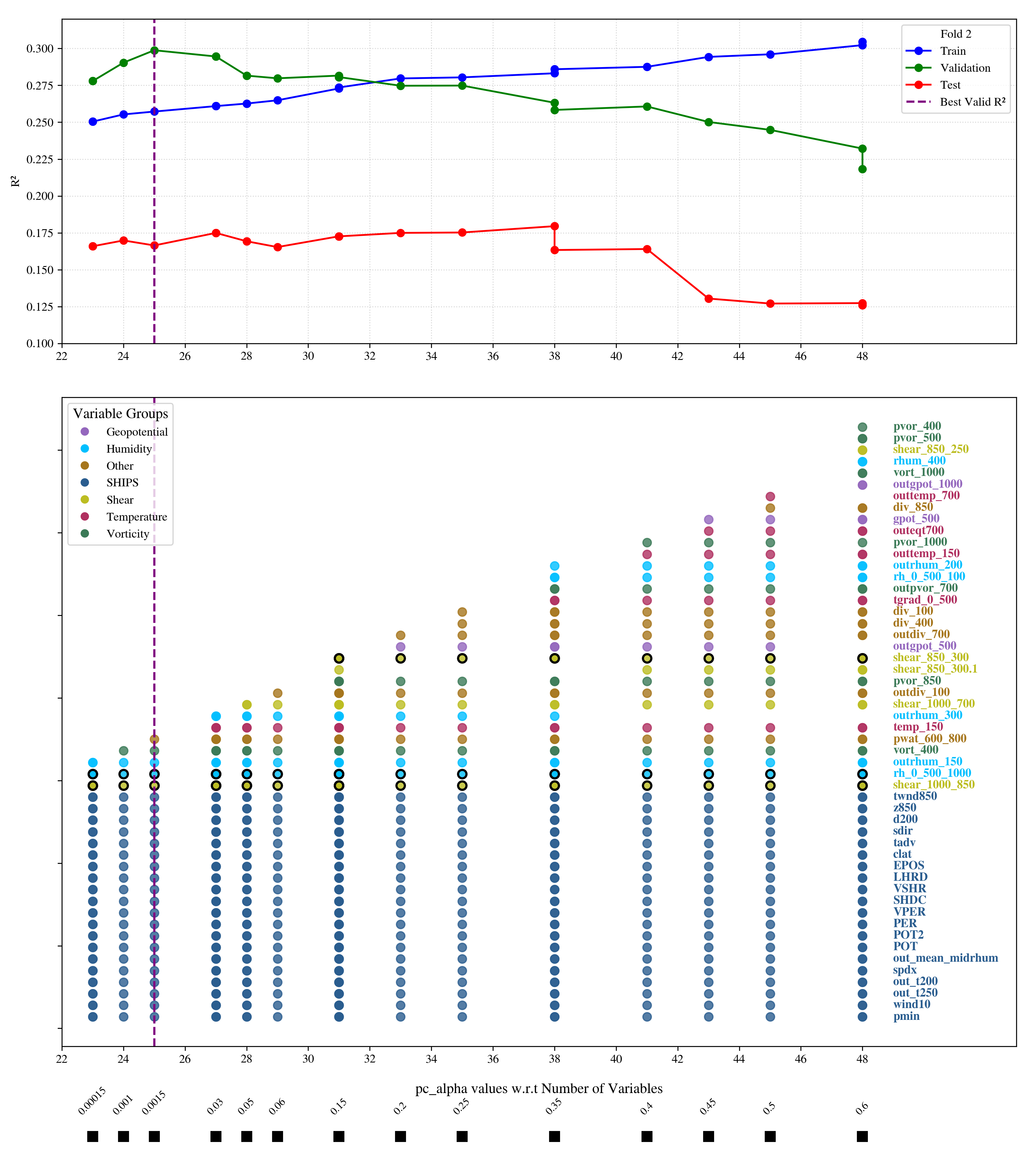}
    \caption{Results for the 72-hour intensity change forecast (DELV72) from the best fold using the SHIPS+ERA5 predictor set \emph{with} SHIPS link assumptions. Same as the Figure S2.
}
    \label{FIGS6}
\end{figure*}

\begin{figure*}[h]
    \centering
    \includegraphics[width=\textwidth]{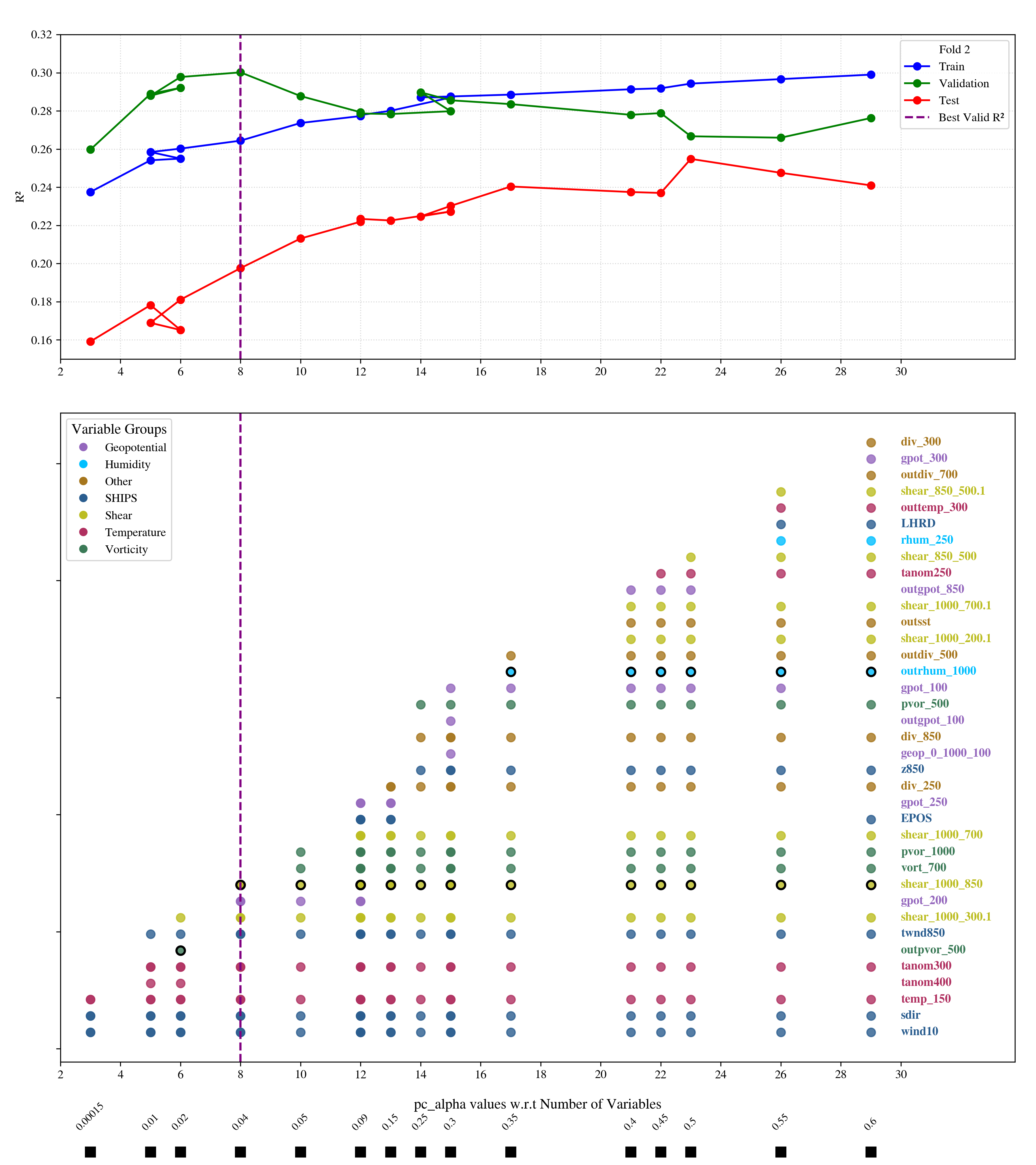}
    \caption{
Results for the 96-hour intensity change forecast (DELV96) from the best fold using the SHIPS+ERA5 predictor set \emph{without} SHIPS link assumptions. Same as the Figure S1.
}
    \label{FIGS7}
\end{figure*}

\begin{figure*}[h]
    \centering
    \includegraphics[width=\textwidth]{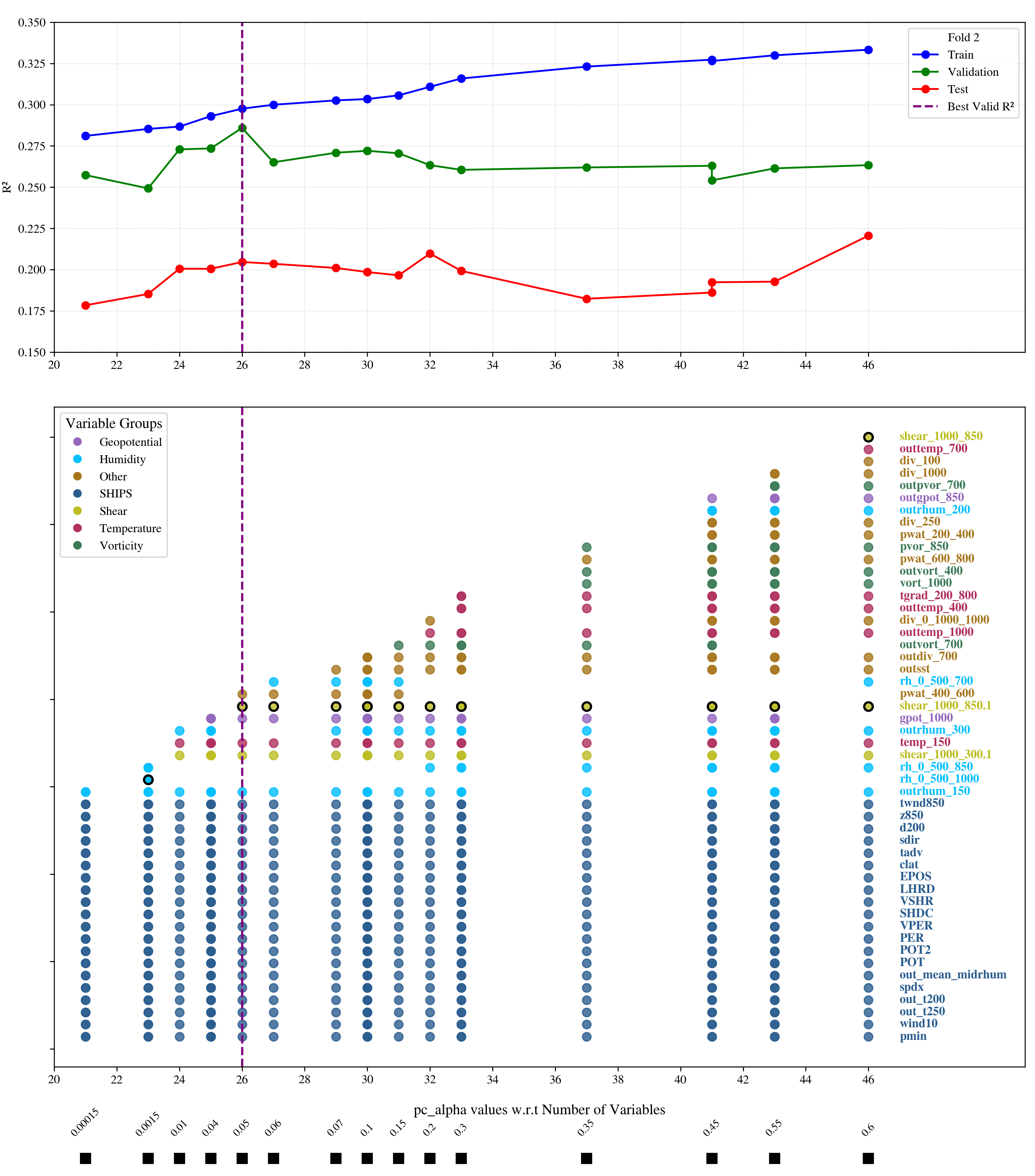}
    \caption{
Results for the 96-hour intensity change forecast (DELV96) from the best fold using the SHIPS+ERA5 predictor set \emph{with} SHIPS link assumptions. Same as the Figure S2.
}
    \label{FIGS8}
\end{figure*}

\begin{figure*}[h]
    \centering
    \includegraphics[width=\textwidth]{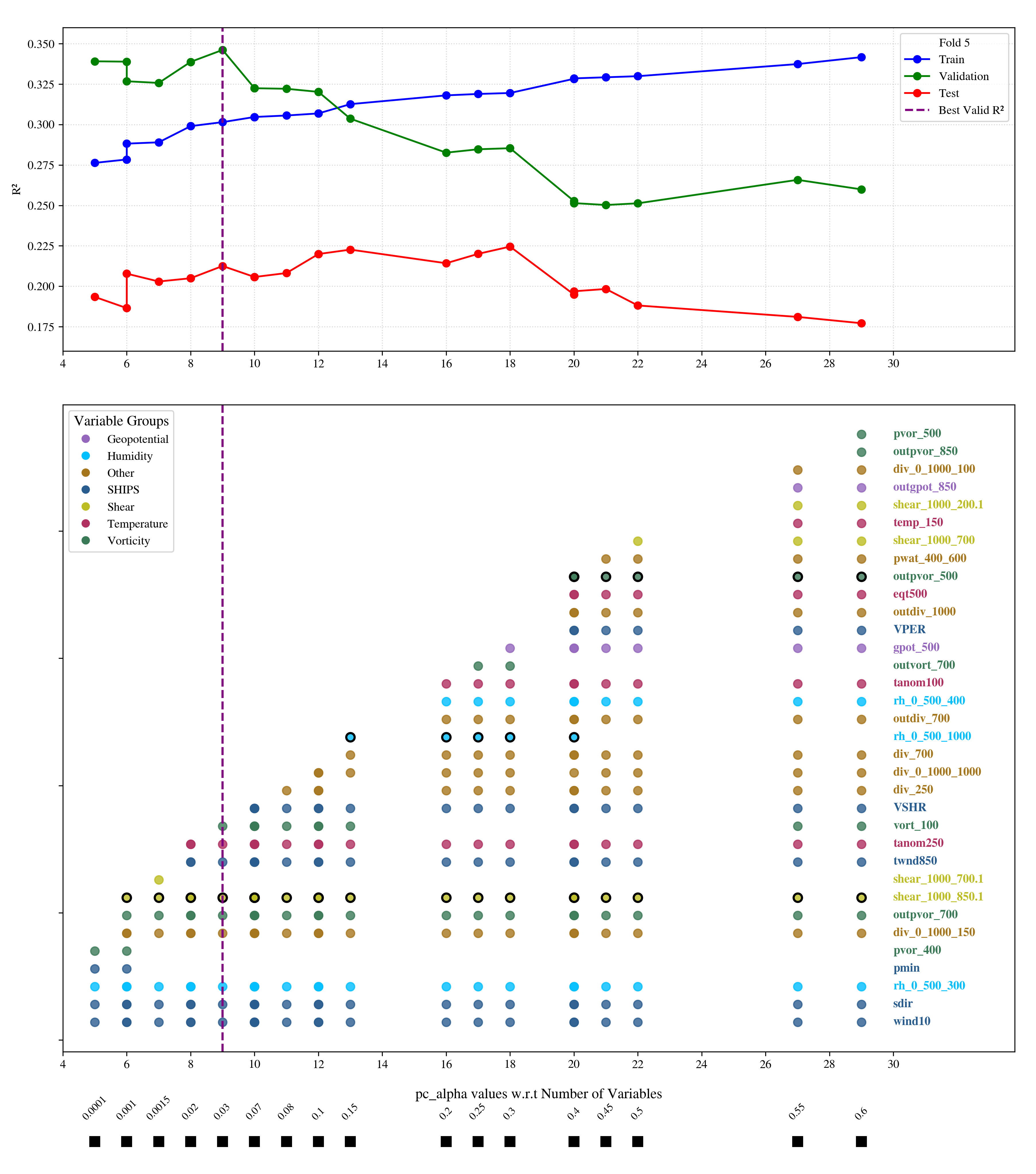}
    \caption{
Results for the 120-hour intensity change forecast (DELV120) from the best fold using the SHIPS+ERA5 predictor set \emph{without} SHIPS link assumptions. Same as the Figure S1.
}
    \label{FIGS9}
\end{figure*}

\begin{figure*}[h]
    \centering
    \includegraphics[width=\textwidth]{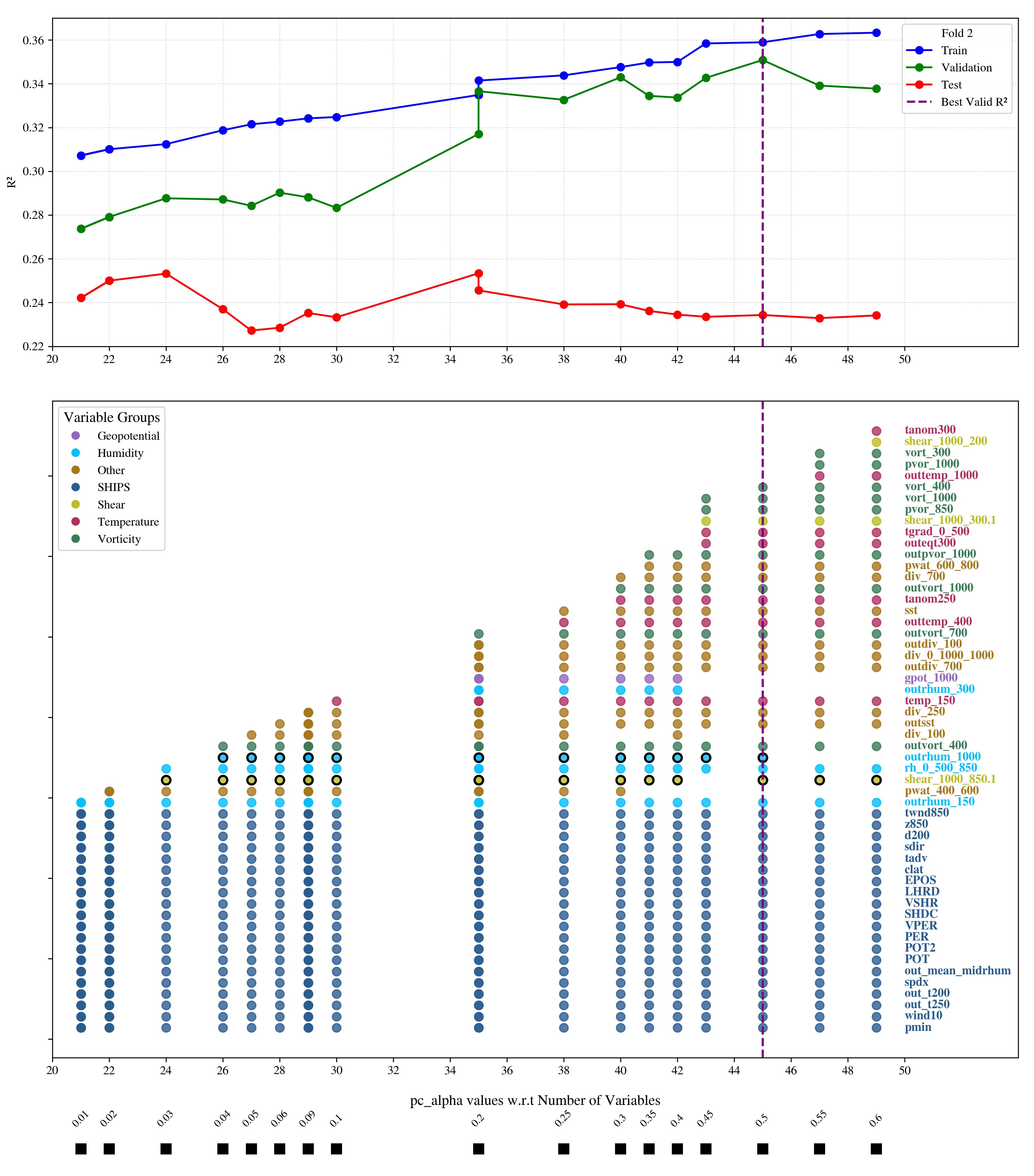}
    \caption{
Results for the 120-hour intensity change forecast (DELV120) from the best fold using the SHIPS+ERA5 predictor set \emph{with} SHIPS link assumptions. Same as the Figure S2.
}
    \label{FIGS10}
\end{figure*}

\begin{figure*}[h]
    \centering
    \includegraphics[height=0.85\textheight, width=1.0\textwidth]{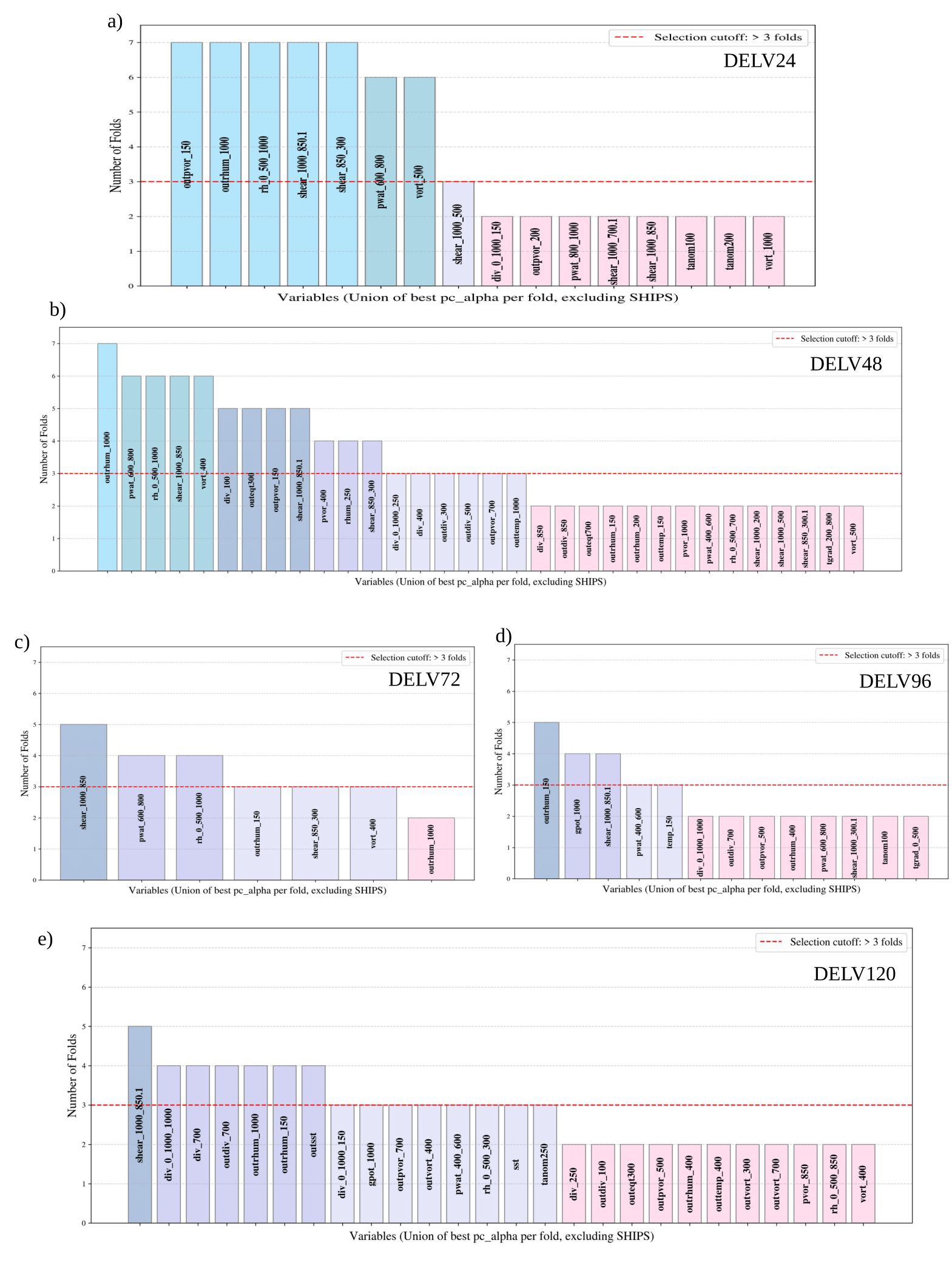}
    \caption{
Bar plots showing the frequency of each variable across the best models from all seven cross-validation folds for experiments with SHIPS Link assumptions for target DELV for lead times 24 hrs to 120 hrs. Red dotted line shows the cut off for variable shortlist.
}
    \label{FIGS11}
\end{figure*}

\begin{figure*}[h]
    \centering
    \includegraphics[height=0.85\textheight, width=1.0\textwidth]{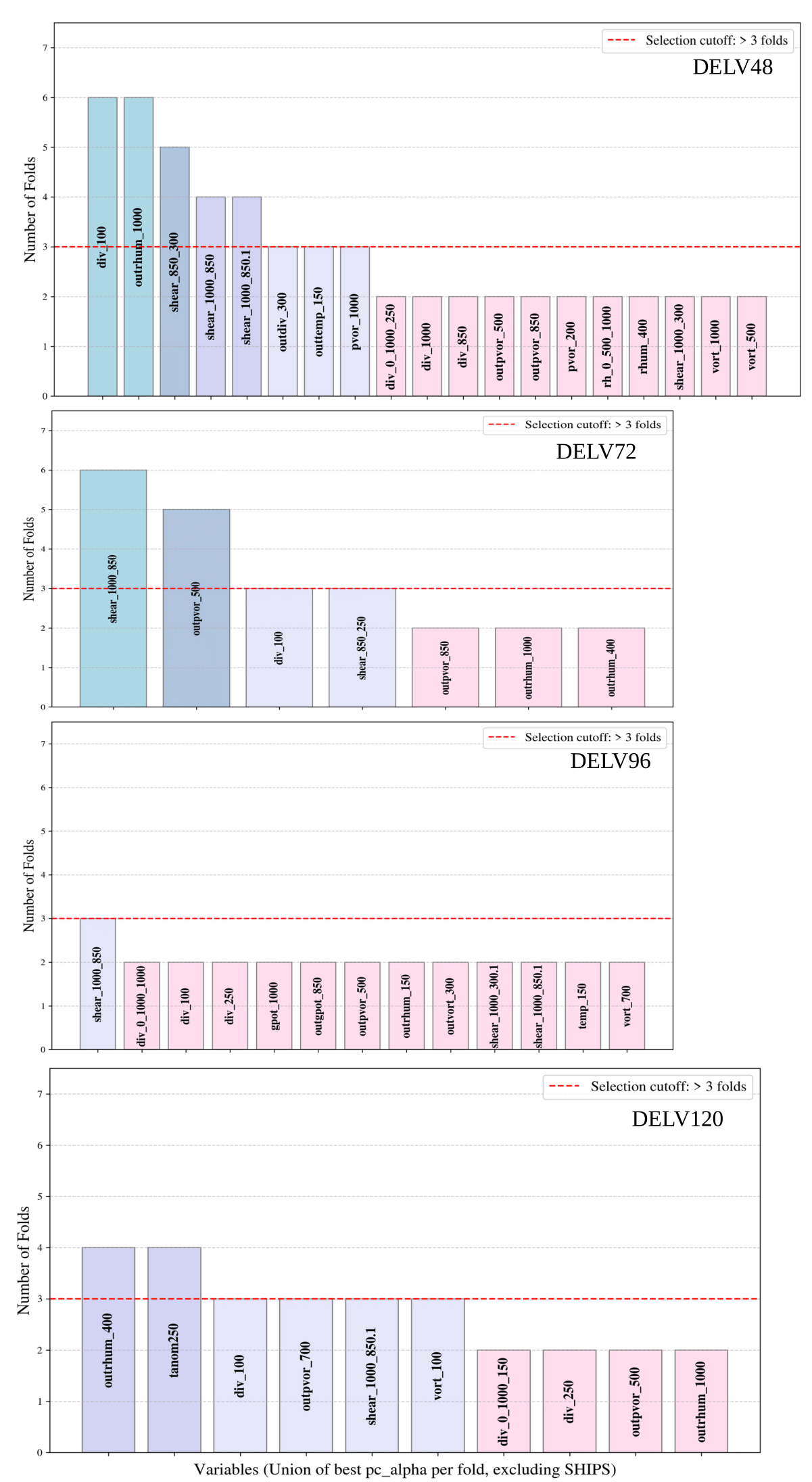}
    \caption{
Bar plots showing the frequency of each variable across the best models from all seven cross-validation folds for experiments without any SHIPS link assumptions for target DELV for lead time 48 hrs to 120 hrs. Red dotted line shows the cut off for variable shortlist.
}
    \label{FIGS12}
\end{figure*}
\begin{figure*}[h]
    \centering
    \includegraphics[height=0.75\textheight, width=1.0\textwidth]{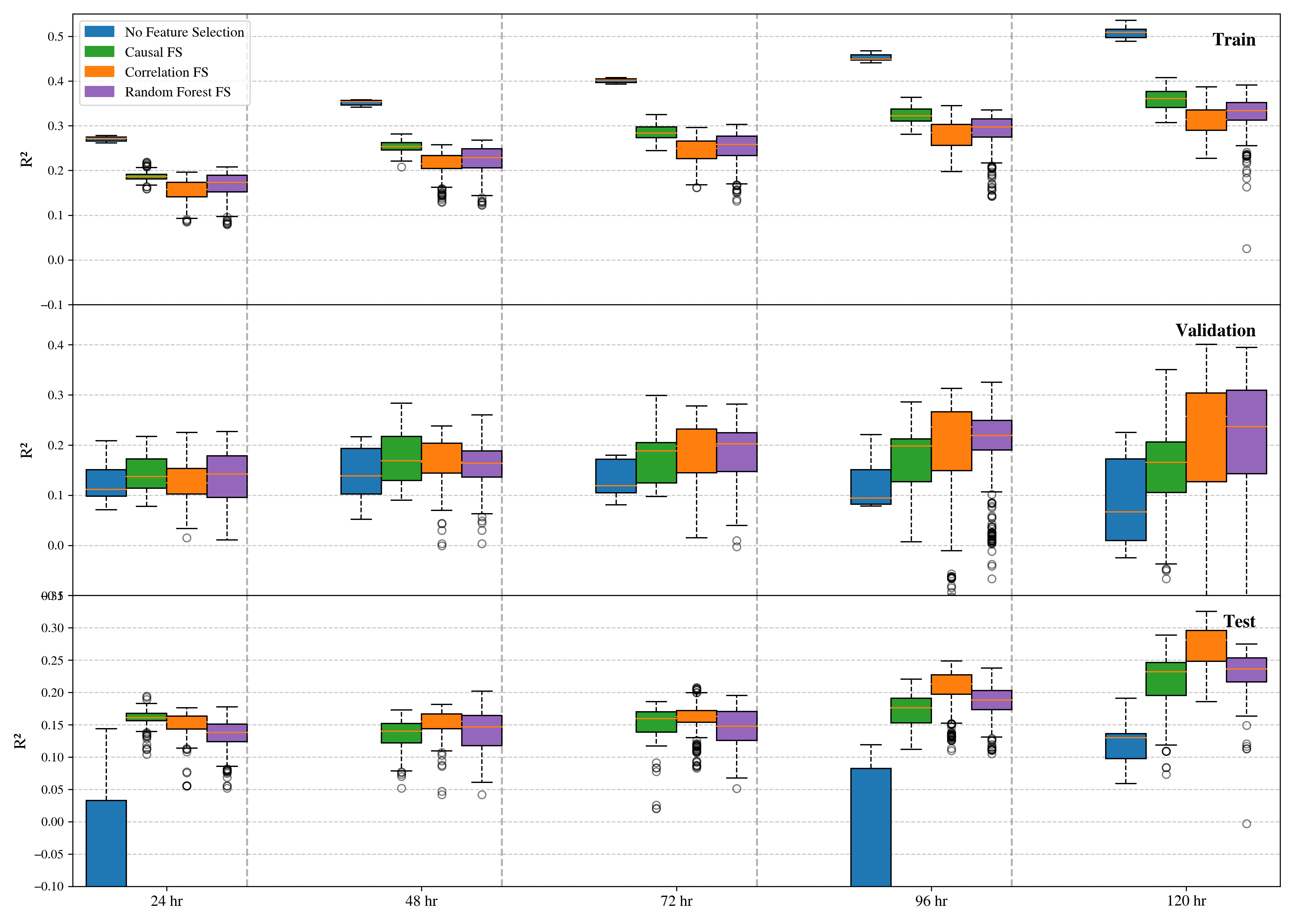}
    \caption{
Boxplot comparing Train (Top), Validation (Middle) and Test (Bottom) R² values for DELV for each lead times 24, 48, 72, 96, 120 hrs for experiments without SHIPS link assumptions similar to Fig.3b, across four feature selection strategies: causal discovery, correlation ranking, random forest importance, and no selection. Causal feature selection yields the highest median R² till 72 hrs lead time, showing improved generalization in a purely statistical prediction setup.
}
    \label{FIGS13}
\end{figure*}

\begin{figure*}[h]
    \centering
    \hspace*{-1cm} 
    \includegraphics[height=0.75\textheight, width=1.05\textwidth]{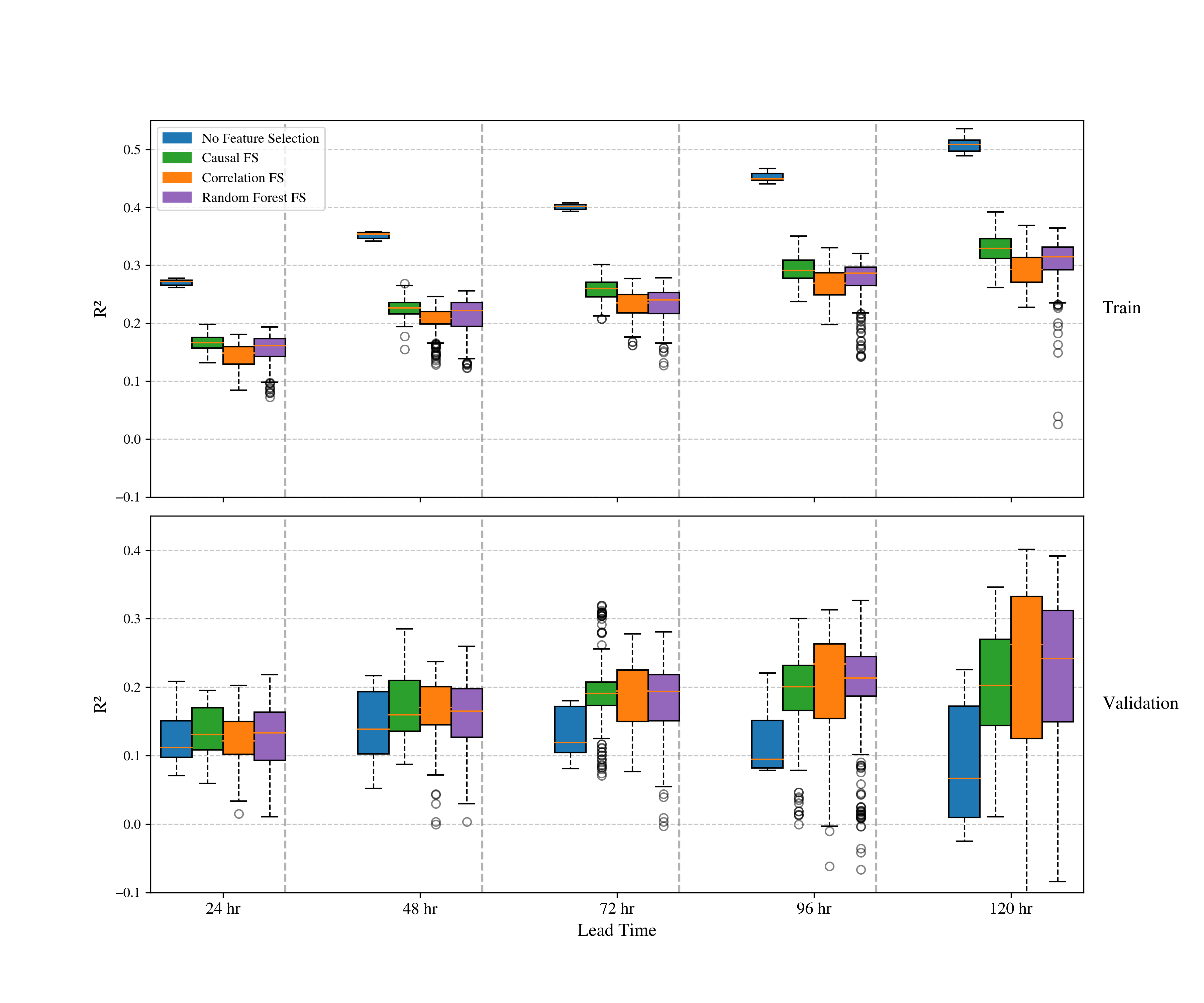}
    \caption{
Boxplot comparing Training (Top), Validation (Bottom) R² values for DELV for each lead times 24, 48, 72, 96, 120 hrs for experiments with SHIPS link assumptions across four feature selection strategies: causal discovery, correlation ranking, random forest importance, and no selection. Causal feature selection yields the highest median R² till 72 hrs lead time, showing improved generalization in a purely statistical prediction setup.
    }
    \label{FIGS14}
\end{figure*}

\begin{figure*}[h]
    \centering
    \hspace*{-1cm} 
    \includegraphics[height=0.75\textheight, width=1.05\textwidth]{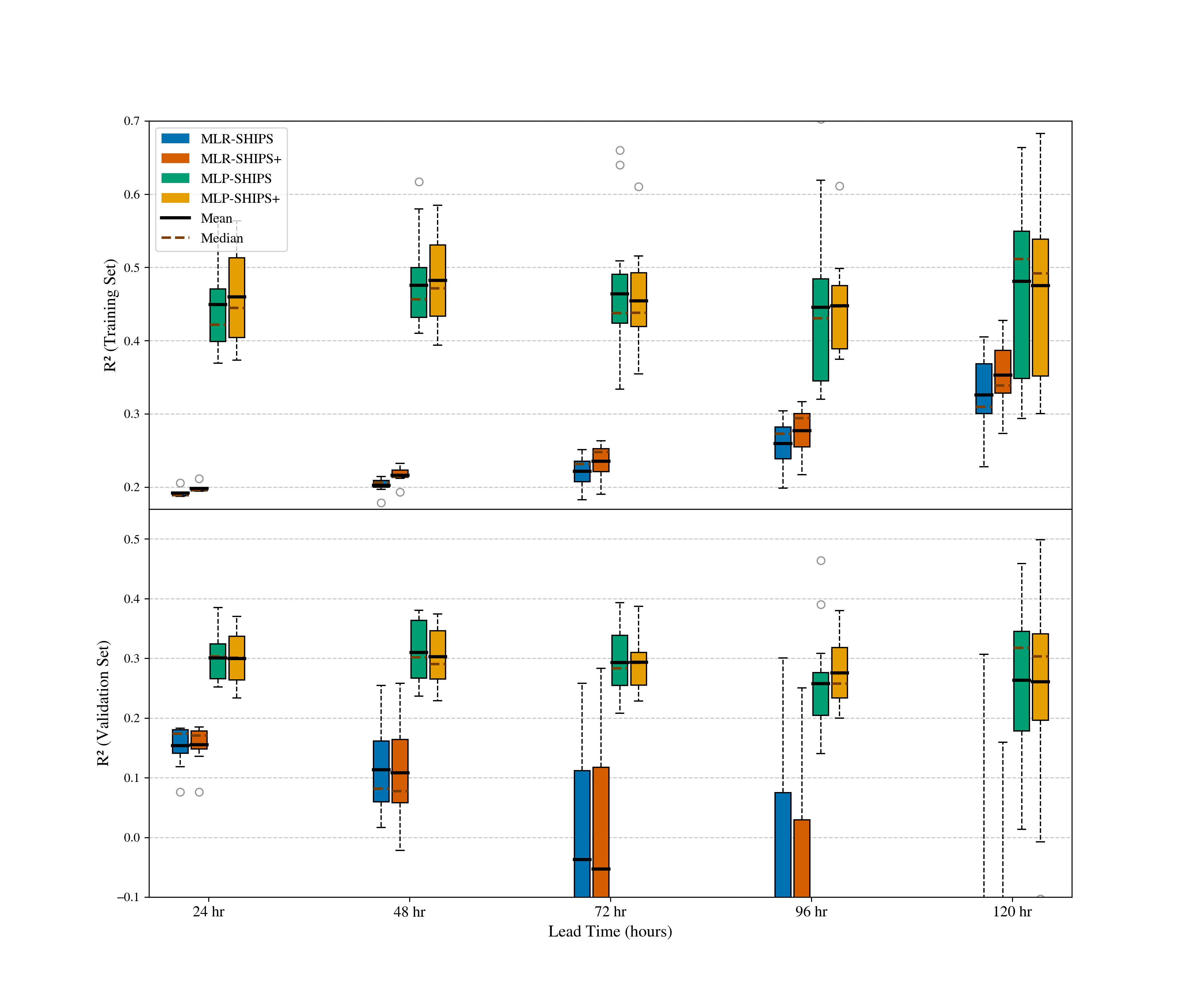}
    \caption{
Boxplot comparing Train (Top), Validation (Bottom) R² values for DELV for each lead times 24, 48, 72, 96, 120 hrs for experiments using SHIPS and SHIPS+ predictors for MLR and MLP. MLP consistently outperforms and have the highest R² values for all lead times showing the nonlinear model's superior ability to capture complex relationships between predictors and intensity change.
    }
    \label{FIGS15}
\end{figure*}



\end{document}